\documentclass[aps,pra,reprint,superscriptaddress,showpacs,showkeys]{revtex4-1}

\usepackage{graphicx}
\usepackage{color}
\usepackage{amssymb}
\usepackage{amsmath}
\usepackage{epsfig}
\usepackage{bm}
\usepackage[american]{babel}
\usepackage[colorlinks,urlcolor=blue,citecolor=blue,linkcolor=blue]{hyperref}

\newcommand{\intraparam}{g_{22}/g_{11}}
\newcommand{\interparam}{\Gamma}
\newcommand{\gtot}{U}
\newcommand{\pseudospinor}{\chi}
\newcommand{\ntot}{{n_\mathrm{tot}}}
\newcommand{\spinvector}{\hat{\mathbf{s}}}
\newcommand{\tcv}[2]{\langle {#1},{#2} \rangle}
\newcommand{\kineticenergy}[1]{T_{#1}}
\newcommand{\trapenergy}[1]{V_{#1}}
\newcommand{\iaenergy}{E_{12}}
\newcommand{\intmeasure}{{\,\mathrm{d}^2 r}}
\newcommand{\windingnumber}{\kappa}
\newcommand{\unitzvector}{\hat{\bf{z}}}

\begin{document}

\title{Ground-state multiquantum vortices in rotating two-species superfluids}
\date{\today}
\author{Pekko~Kuopanportti}\email{pekko.kuopanportti@monash.edu}
\affiliation{School of Physics and Astronomy, Monash University, Victoria 3800, Australia}
\affiliation{Department of Physics, University of Helsinki, P.O. Box 43, FI-00014 Helsinki, Finland}
\author{Natalia~V.~Orlova}
\affiliation{Departement Fysica, Universiteit Antwerpen, Groenenborgerlaan 171, B-2020 Antwerpen, Belgium}
\author{Milorad~V.~Milo\v{s}evi\'{c}}\email{milorad.milosevic@uantwerpen.be}
\affiliation{Departement Fysica, Universiteit Antwerpen, Groenenborgerlaan 171, B-2020 Antwerpen, Belgium}

\begin{abstract}
We show numerically that a rotating, harmonically trapped mixture of two Bose--Einstein-condensed superfluids can---contrary to its single-species counterpart---contain a multiply quantized vortex in the ground state of the system. This giant vortex can occur without any accompanying single-quantum vortices, may either be coreless or have an empty core, and can be realized in a ${}^{87}$Rb--${}^{41}$K Bose--Einstein condensate. Our results not only provide a rare example of a stable, solitary multiquantum vortex but also reveal exotic physics stemming from the coexistence of multiple, compositionally distinct condensates in one system.
\end{abstract}

\maketitle

\section{Introduction}

According to the conventional paradigm~\cite{Ons1949.NCim6Sup2.249,Fey1955.PLTP1.17}, the ground state in a rotating superfluid will involve only singly quantized vortices (SQVs). Vortices with larger quantum numbers are energetically unfavorable and do not occur---not even for rapid rotation, which instead spawns a triangular Abrikosov lattice of SQVs~\cite{Abr1957.ZETF32.1442}. Although this is well established~\cite{Yar1979.PRL43.214,Abo2001.Sci292.476,Zwi2005.Nat435.1047} for a solitary superfluid described by a single $\mathbb{C}$-valued order parameter $\Psi$, vortex physics becomes much more diverse when \emph{multiple} mutually interacting superfluids are rotated simultaneously in the same container. 

Already for the simplest mixture, which consists of two superfluid species and is described by two $\mathbb{C}$-valued order parameters $\Psi_1$ and $\Psi_2$, a myriad of unusual ground-state vortex structures have been found in experimental and theoretical studies~\cite{Kas2005.IJMPB19.1835}. Experimentally, a versatile platform to study vortices is provided by atomic Bose--Einstein condensates~(BECs)~\cite{Mad2000.PRL84.806,And2010.JLTP161.574,Mat1999.PRL83.2498}, in which two-component superfluid mixtures have been realized using two different spin states of the same isotope~\cite{Sch2004.PRL93.210403,Mya1997.PRL78.586,Hal1998.PRL81.1539,Mat1999.PRL83.2498,Del2001.PRA63.051602,And2009.PRA80.023603}, two different isotopes of the same element~\cite{Bap2008.PRL101.040402,Sug2011.PRA84.011610}, or two distinct elements~\cite{Fer2002.PRL89.053202,Mod2002.PRL89.190404,Cat2008.PRA77.011603,Tha2008.PRL100.210402,Aik2009.NJP11.055035,Cat2009.PRL103.140401,Car2011.PRA84.011603,Ler2011.EPJD65.3,Pas2013.PRA88.023601}. The unconventional vortex structures that were detected in these experiments comprise coreless SQVs~\cite{Mat1999.PRL83.2498} and square vortex lattices~\cite{Sch2004.PRL93.210403}. Theoretical studies, however, have furnished the two-species BECs with many more ground-state vortex configurations than the aforementioned two~\cite{Kas2004.PRL93.250406,Kas2005.PRA71.043611,Mas2011.PRA84.033611,Mue2002.PRL88.180403,Kas2003.PRL91.150406,Kec2006.PRA73.023611,Min2009.PRA79.013605,Dah2008.PRB78.144510,Kuo2012.PRA85.043613}: Predicted but hitherto unobserved ones include serpentine vortex sheets~\cite{Kas2009.PRA79.023606}, triangular lattices of vortex pairs~\cite{Kuo2012.PRA85.043613}, and, in a pseudospin-$1/2$ representation, giant skyrmions~\cite{Yan2008.PRA77.033621,Mas2011.PRA84.033611} and meron pairs~\cite{Kas2004.PRL93.250406,Kas2005.PRA71.043611}.

One peculiar feature of the two-species mixture, which goes against the traditional paradigm, is the appearance of multiply quantized vortices~(MQVs) in the rotating ground state of the harmonically trapped system~\cite{Yan2008.PRA77.033621,Mas2011.PRA84.033611,Kuo2012.PRA85.043613}. So far, the MQVs, also known as giant vortices, have been predicted only in complicated states involving a number of accompanying SQVs and a large total circulation, thereby requiring rotation frequencies close to the maximum set by the harmonic trap frequency. Consequently, the states have eluded experimental observation and, due to the accompanying SQVs, might not be suitable for investigating the rarely encountered ground-state MQV in a controlled fashion. Besides being exotic and interesting in their own right, MQVs could also be used to realize bosonic quantum Hall states~\cite{Ron2011.SciRep1.43}, initiate quantum turbulence~\cite{Abr1995.PRB52.7018,Ara1996.PRB53.75,San2014.arXiv.1405.0992}, or implement a ballistic quantum switch~\cite{Mel2002.Nat415.60}.

In this article, we make the ground-state MQVs more accessible to experiments by showing theoretically that an interacting mixture of two dilute superfluids, when rotated at moderate speed, exhibits ground states that contain a \emph{solitary} MQV in one of the superfluids. We find such states both for mutually attractive mixtures, where the MQV has a completely empty core, and for mutually repulsive mixtures, where the core is occupied by the other, vortex-free superfluid species. These states represent a rare instance of a stable, solitary MQV in an atomic BEC and, as such, constitute a robust, well-isolated, and tunable environment for the experimental exploration of MQV physics, in complement to earlier observations in mesoscopic superconductors~\cite{Sch1998.PRL81.2783,Kan2004.PRL93.257002,Gri2007.PRL99.147003,Cre2011.PRL107.097202}.

All the discovered states share the property that the two superfluid species carry unequal numbers of circulation quanta under the same external rotation. This requires the two superfluids to be composed of particles with sufficiently different masses~\cite{note_circulation}. For concreteness, we will focus on the harmonically trapped two-species BEC of $^{87}$Rb and ${}^{41}$K because it has already been realized in several experiments~\cite{Fer2002.PRL89.053202,Mod2002.PRL89.190404,Cat2008.PRA77.011603,Tha2008.PRL100.210402,Aik2009.NJP11.055035,Cat2009.PRL103.140401}, it enables a flexible control over its interaction strengths~\cite{Tha2008.PRL100.210402}, and it has a suitable atomic mass ratio of $\sim\!\!2$. Although we present ground states only for this particular system, the essential features of our results apply generally to mass-imbalanced binary mixtures of dilute superfluids. 

\section{Model}

We assume that the two-species BEC is rotated with angular velocity $\Omega\unitzvector$. In the zero-temperature mean-field regime, the ground-state order parameters $\Psi_1$ (assigned to ${}^{87}$Rb) and $\Psi_2$ (${}^{41}$K) satisfy the coupled time-independent Gross--Pitaevskii equations in the rotating reference frame~\cite{Kas2005.IJMPB19.1835}:
\begin{equation}
\left( {\cal H}_j + g_{jj} |\Psi_j|^2 + g_{12}|\Psi_{3-j}|^2 - \mu_j \right) \Psi_j\left(r,\phi\right) =  0,\label{eq:GPE}
\end{equation}
where $j\in\left\{1,2\right\}$, 
\begin{equation}
{\cal H}_j = -\frac{\hbar^2}{2m_j}\nabla^2+\frac{1}{2} m_j\omega_j^2 r^2+i\hbar\Omega\frac{\partial}{\partial\phi},
\end{equation}
and the chemical potentials $\mu_j$ ensure that $\int\! |\Psi_j|^2 \intmeasure =N_j$. Here $N_j$, $m_j$, and $\omega_j$ denote, respectively, the total number, the mass, and the radial harmonic trapping frequency of species $j$ atoms. We only consider quasi-two-dimensional configurations pertaining to, e.g., highly oblate (prolate) traps with strong (weak) axial confinement and $\Psi_j$ approximately Gaussian (constant) in the axial direction. The intraspecies interaction strengths $g_{jj}$ are assumed to be positive, whereas for the interspecies parameter $g_{12}$ we also consider negative values. 

We parametrize the interactions by the three dimensionless quantities $\gtot = \left(g_{11} + g_{12}\right)m_1 N_1/\hbar^2$, $\intraparam$, and $\interparam= g_{12}/\sqrt{g_{11} g_{22}}> -1$. The ground states, i.e., the lowest-energy solutions of Eqs.~\eqref{eq:GPE}, are then uniquely specified by these three and the following four other parameters: $m_2/m_1$, $N_2/N_1$,  $\omega_2^2/\omega_1^2$, and $\Omega/\omega_1$. Focusing on the ${}^{87}$Rb--${}^{41}$K BEC, we fix $m_2/m_1=0.471$. Equations~\eqref{eq:GPE} are solved numerically using link-variable discretization~\cite{Geu2008.PRA78.053610} and gradient descent. 

\section{Ground-state multiquantum vortices}

In order to understand why MQVs emerge in the rotating two-species BEC, we begin with a scenario where only $\interparam$ is varied while the other parameters are held constant. Furthermore, for $\windingnumber_j \in \mathbb{Z}$, let $\tcv{\windingnumber_1}{\windingnumber_2}$ denote a sufficiently pointlike phase defect about which $\mathrm{arg}\left(\Psi_1\right)$ winds by $\windingnumber_1 \times 2 \pi$ and $\mathrm{arg}\left(\Psi_2\right)$ by $\windingnumber_2 \times 2 \pi$. For interspecies repulsion ($\interparam > 0$), the simplest MQV state appearing as the ground state is a $\tcv{2}{0}$ vortex, whereas for $\interparam < 0$, the simplest one corresponds to $\tcv{2}{1}$. Below, we investigate these two cases separately.

\begin{figure}
\includegraphics[width=0.8\columnwidth,keepaspectratio]{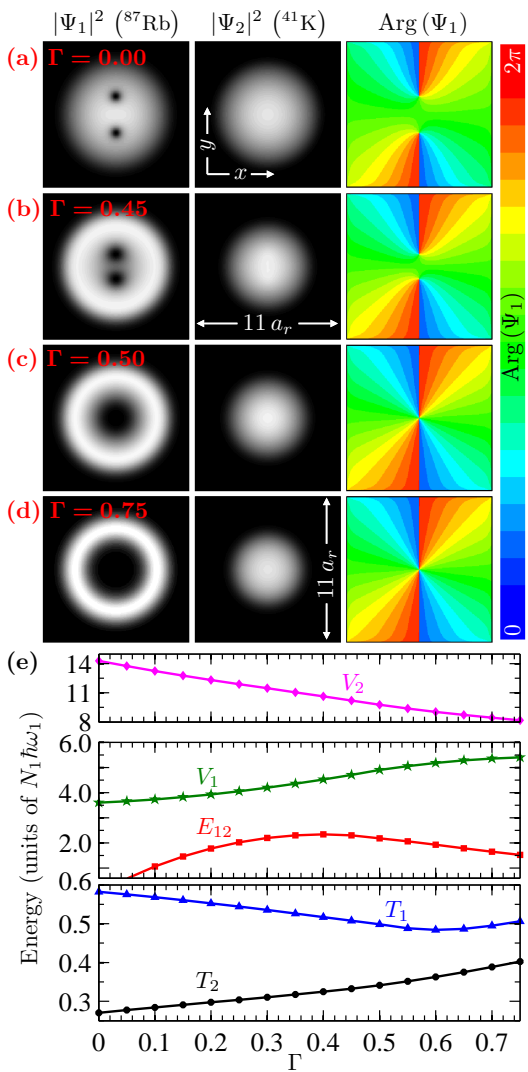}
\caption{\label{fig:2-0-vortex-profiles} Emergence of a ground-state two-quantum vortex in a rotating, mutually repulsive two-species ${}^{87}$Rb--${}^{41}$K BEC. (a)--(d) Atomic densities $|\Psi_1|^2$ and $|\Psi_2|^2$ and the complex phase of the order parameter $\Psi_1$ in the ground state at the indicated value of the interspecies interaction strength $\interparam = g_{12}/\sqrt{g_{11} g_{22}}$. In all four cases, $\mathrm{Arg}\left(\Psi_2\right)\equiv \textrm{const}$~(not shown). (e)~Interspecies interaction energy $\iaenergy = g_{12} \int\! |\Psi_1 \Psi_2|^2 \intmeasure$, kinetic energies $\kineticenergy{j}=  \hbar^2 \int\! |\nabla \Psi_j|^2 \intmeasure/2 m_j$, and trap energies $\trapenergy{j}= m_j \omega_j^2 \int\! r^2 |\Psi_j|^2  \intmeasure/2$ as functions of $\interparam$ for the same ground states. The parameters in the Gross--Pitaevskii equations are set to $m_2/m_1=0.471$, $\intraparam=4$, $\omega_2^2/\omega_1^2 = 10$, $N_2/N_1=1$, $\Omega/\omega_1=0.4$, and $\gtot = \left(g_{11} + g_{12}\right)m_1 N_1 /\hbar^2 = 300$. The length unit is $a_r=\sqrt{\hbar/m_1\omega_1}$.}
\end{figure}

Consider first the mutually repulsive mixture. Figures~\ref{fig:2-0-vortex-profiles}(a)--\ref{fig:2-0-vortex-profiles}(d) depict ground states at different $\interparam\geq 0$ for a rotating ${}^{87}$Rb--${}^{41}$K BEC in which there are two circulation quanta in Rb and none in K. Figure~\ref{fig:2-0-vortex-profiles}(e) shows the relevant energy terms as a function of $\interparam$. When $\interparam=0$ [Fig.~\ref{fig:2-0-vortex-profiles}(a)], the two off-centered $\tcv{1}{0}$ vortices are separated by a distance of $\sim\!\!10$ times their core radius. As $\interparam$ increases, the condensates move apart, with Rb shifting outward and K inward; this behavior is manifested in the trap potential energy, which increases for Rb and decreases for K. Consequently, Rb is depleted from the region between the two $\tcv{1}{0}$ vortices, enabling them to merge into a $\tcv{2}{0}$ vortex without the kinetic-energy increase typical of MQV formation; indeed, the kinetic energy $\kineticenergy{1}$ of Rb decreases with $\interparam\in\left[0,0.6\right]$. Hence, for $\interparam \geq 0.5$, we observe an axisymmetric $\tcv{2}{0}$ vortex, about which $\mathrm{arg}\left(\Psi_1\right)$ winds by $2\times 2\pi$. It is a coreless vortex~\cite{Kas2004.PRL93.250406,Kas2005.PRA71.043611,Mas2011.PRA84.033611,Ho1978.PRB18.1144,Miz2002.PRL89.030401,Miz2002.PRA66.053610,Miz2004.PRA70.043613} in the sense that the total atomic density $\ntot= |\Psi_1|^2+|\Psi_2|^2$ does not vanish at the phase singularity. We stress that the $\tcv{2}{0}$ vortex is a unique example of a ground-state MQV in a purely harmonic trap that occurs as a solitary topological defect without any accompanying SQVs. 

\begin{figure}
\includegraphics[width=0.8\columnwidth,keepaspectratio]{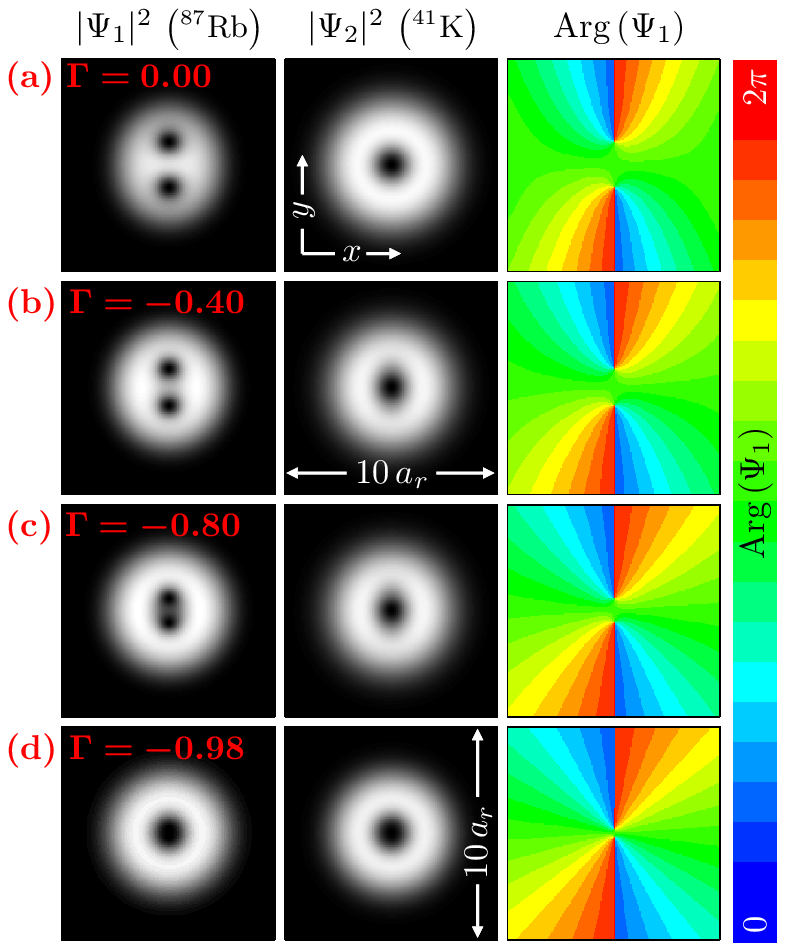}
\caption{\label{fig:2-1-vortex-profiles} Formation of a ground-state vortex with specieswise quantum numbers $\tcv{\windingnumber_1}{\windingnumber_2}=\tcv{2}{1}$ in a rotating, mutually attractive ${}^{87}$Rb--${}^{41}$K BEC. (a)--(d) Atomic densities $|\Psi_1|^2$ and $|\Psi_2|^2$ and the complex phase of $\Psi_1$ in the ground state at the indicated value of $\interparam = g_{12}/\sqrt{g_{11} g_{22}}$; $\mathrm{Arg}\left[ \Psi_2 \left(r,\phi\right)\right] \simeq \phi$ for all $\interparam$~(not shown). Here $\intraparam=\omega_2^2/\omega_1^2 =N_2/N_1=1$, $\Omega/\omega_1=0.7$, and $\gtot  = 50$. }
\end{figure}

The emergence of the ground-state $\tcv{2}{1}$ vortex for $\interparam < 0$ is illustrated in Fig.~\ref{fig:2-1-vortex-profiles}. For uncoupled condensates [Fig.~\ref{fig:2-1-vortex-profiles}(a)], there are two off-centered $\tcv{1}{0}$ vortices and one central $\tcv{0}{1}$ vortex. As $\interparam$ approaches $-1$, the two $\tcv{1}{0}$ vortices move closer to each other, so that at $\interparam = -0.98$, all three phase singularities lie at the origin and make up an axisymmetric $\tcv{2}{1}$ vortex. To explain the movement, we note that the kinetic energy $\kineticenergy{1}$ increases when the two vortices approach each other, whereas the interspecies interaction energy $\iaenergy$ decreases due to the increasing overlap $\int\! n_1 n_2 \intmeasure$. Since $\iaenergy$ gains in importance when the attraction becomes stronger, it eventually begins to dominate over $\kineticenergy{1}$, and thus the $\tcv{2}{1}$ vortex forms~[Fig.~\ref{fig:2-1-vortex-profiles}(d)].

The ground-state $\tcv{2}{1}$ vortex constitutes a rare instance of a stable MQV with a genuinely empty, self-supporting core. Typically, such vortices are rendered unstable against splitting by quasiparticle excitations that are highly localized within the core~\cite{Dod1997.PRA56.587,Rok1997.PRL79.2164,Pu1999.PRA59.1533,Iso1999.PRA60.3313,Svi2000.PRL84.5919,Vir2001.PRL86.2704,Kaw2004.PRA70.043610,Jac2005.PRA72.053617,Huh2006.PRA74.063619,Lun2006.PRA74.063620,Cap2009.JPB42.145301,Kuo2010.PRA81.023603,Kuo2010.PRA81.033627}. In our case, however, the doubly quantized vortex in $\Psi_1$ is held together by the indivisible SQV in $\Psi_2$.

The $\tcv{2}{1}$ vortices are most readily found for relatively small values of $\gtot$. This is because small $\gtot$ implies a large size of the vortex cores, which suppresses the kinetic energy near the phase singularities and leads to strong dependence of $\iaenergy$ on the vortex positions. 

To produce the MQVs of Figs.~\ref{fig:2-0-vortex-profiles} and \ref{fig:2-1-vortex-profiles}, it is desirable to have control over the parameter $\interparam=g_{12}/\sqrt{g_{11} g_{22}}$. 
In experiments, $g_{jk}$ may be tuned with Feshbach resonances~\cite{Chi2010.RMP82.1225}, which have been demonstrated for ${}^{87}$Rb--${}^{87}$Rb~\cite{Mar2002.PRL89.283202,Er2004.PRA69.032705,Wid2004.PRL92.160406}, ${}^{41}$K--${}^{41}$K~\cite{Err2007.NJP9.223,Kis2009.PRA79.031602}, and ${}^{87}$Rb--${}^{41}$K~\cite{Tha2008.PRL100.210402} interactions. However, ground-state MQVs can also be obtained in the ${}^{87}$Rb--${}^{41}$K BEC without employing Feshbach resonances. To demonstrate this for an axially uniform system, we use the bare $s$-wave scattering lengths $a_{11}/a_{\mathrm{B}}=99$~\cite{Kem2002.PRL88.093201}, $a_{22}/a_{\mathrm{B}}= 60$~\cite{Wan2000.PRA62.052704}, and $a_{12}/a_{\mathrm{B}} = 163$~\cite{Fer2002.PRL89.053202}, where $a_{\mathrm{B}}$ is the Bohr radius, and accordingly set $\intraparam = 1.29$ and $\interparam = 2.27$. The remaining parameters are fixed after the ${}^{87}$Rb--${}^{41}$K experiment of Ref.~\cite{Cat2009.PRL103.140401}. 

\begin{figure}
\includegraphics[width=0.8\columnwidth,keepaspectratio]{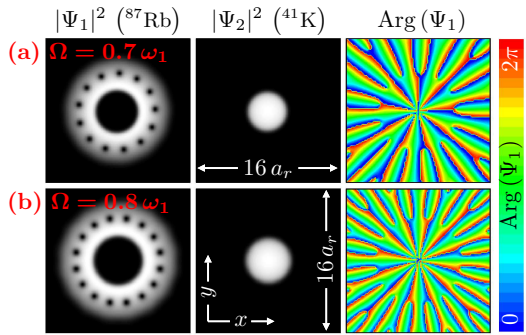}
\caption{\label{fig:skyrmion-profiles} Ground states of a ${}^{87}$Rb--${}^{41}$K BEC with interaction parameters $\interparam=2.27$ and $\intraparam=1.29$ corresponding to the unmodified scattering lengths in a highly prolate trap, shown for two different rotation frequencies. In both states, $\mathrm{Arg}\left(\Psi_2\right)\equiv \textrm{const}$~(not shown). Furthermore, $\omega_2^2/\omega_1^2=2.12$, $N_2/N_1=0.27$, and $\gtot=2800$ after the experiment of Ref.~\cite{Cat2009.PRL103.140401}.}
\end{figure}

Figure~\ref{fig:skyrmion-profiles} shows the resulting ground states at two different rotation frequencies, $\Omega/\omega_1 = 0.7$ and $0.8$. At $\Omega/\omega_1 = 0.7$ (0.8), the Rb species hosts a central 9-quantum (12-quantum) giant vortex surrounded by a ring of 13 (16) SQVs. In both cases, the K species is vortex-free and occupies the core of the central giant vortex. At larger $\Omega$, the giant vortex becomes surrounded by a triangular lattice of SQVs; similar profiles have been found earlier for rapidly rotating single-component BECs in anharmonic trap potentials~\cite{Fis2003.PRL90.140402,Jos2004.Chaos14.875,Aft2004.PRA69.033608,Jac2004.PRA69.053619,Fet2005.PRA71.013605,Kim2005.PRA72.023619,Fu2006.PRA73.013614,Cor2007.JMP48.042104,Cor2011.PRA84.053614}.

\section{Pseudospin textures}

In this section, we analyze our results in the pseudospin-$1/2$ representation~\cite{Mat1999.PRL83.3358,Mue2004.PRA69.033606,Kas2004.PRL93.250406,Kas2005.PRA71.043611,Kas2005.IJMPB19.1835}. At points where $\ntot\neq 0$, we define the local unit-length pseudospin 
\begin{equation}
\spinvector\left(r,\phi\right)=\frac{1}{\ntot\left(r,\phi\right)}\sum_{jk} \Psi_j^\ast\left(r,\phi\right) \bm{\sigma}_{jk} \Psi_k\left(r,\phi\right),
\end{equation} 
where $\bm{\sigma}$ is a vector of the three Pauli matrices. Now consider a $\tcv{\windingnumber_1}{\windingnumber_2}$ vortex about which the atomic densities are locally axisymmetric. After transforming to shifted polar coordinates $\left(r',\phi'\right)$ with the vortex core at $r'=0$, we can write $\Psi_j$, for small $r'$, in terms of a spin rotation $Z$ and a $\mathrm{U}\left(1\right)$ gauge transformation acting on a unit-length reference spinor $\pseudospinor\in\mathbb{C}^2$: 
\begin{eqnarray}
\Psi_j\left(r',\phi'\right) &=& |\Psi_j\left(r'\right)|  e^{i\left( \windingnumber_j \phi'+ C_j \right)}  \\ &=& \sqrt{\ntot \left(r'\right) } e^{\frac{i}{2}\windingnumber_\mathrm{g}\phi'} \sum_{k}  Z_{jk}\left( \windingnumber_\mathrm{s} \phi' \right) \pseudospinor_{k}\left(r'\right),\nonumber
\end{eqnarray}
where $\windingnumber_\mathrm{s}=\windingnumber_2-\windingnumber_1$ and $\windingnumber_\mathrm{g}=\windingnumber_1+\windingnumber_2$ are integers that determine, respectively, the number of $2\pi$ rotations of $\spinvector$ about the unit vector $\unitzvector$ and the number of $\pi$ windings of the $\mathrm{U}\left(1\right)$ gauge along a contour enclosing the core, $Z\left( \windingnumber_\mathrm{s} \phi' \right) = \exp\left(-i \windingnumber_\mathrm{s} \phi' \sigma_z / 2 \right)$, and $C_j\in\mathbb{R}$ are constants.

\begin{figure}
\includegraphics[width=0.8\columnwidth,keepaspectratio]{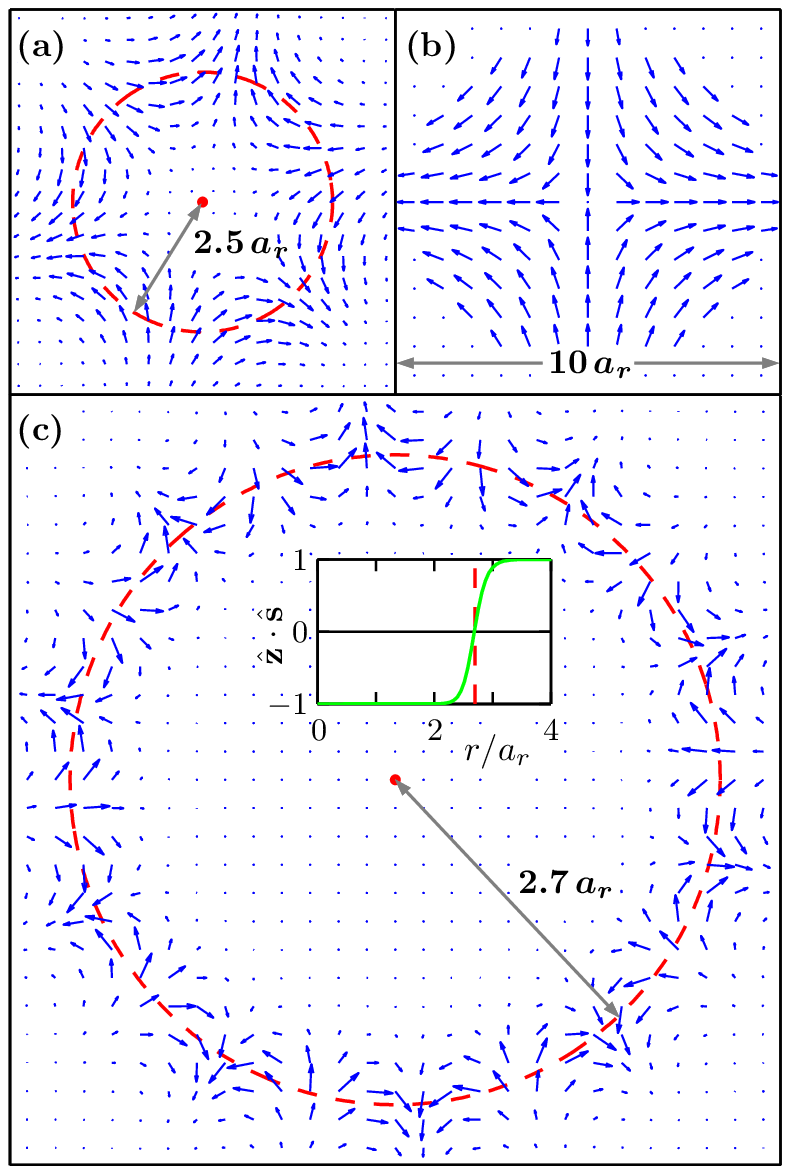}
\caption{\label{fig:pseudospin-textures} Pseudospin textures of a (a)~two-quantum skyrmion~[for the state in Fig.~\ref{fig:2-0-vortex-profiles}(d)], (b)~single-quantum spin vortex~[Fig.~\ref{fig:2-1-vortex-profiles}(d)], and (c)~nine-quantum giant skyrmion~[Fig.~\ref{fig:skyrmion-profiles}(a)]. The arrows represent the projection of the local pseudospin $\spinvector=\sum_{jk}\Psi_j^\ast{\bm{\sigma}}_{jk} \Psi_k/\ntot$ onto the $xy$ plane. Here $\bm{\sigma}$ is a vector of the Pauli matrices and $\ntot=|\Psi_1|^2+|\Psi_2|^2$. The dashed circle is the species interface, where $|\Psi_1|=|\Psi_2|$. The inset in (c) shows the $z$ projection of $\spinvector$ as a function of the radial coordinate $r$.}
\end{figure}

Figure~\ref{fig:pseudospin-textures} shows $\spinvector$ for some of the ground states in Figs.~\ref{fig:2-0-vortex-profiles}--\ref{fig:skyrmion-profiles}. The $\tcv{2}{0}$ vortex in Fig.~\ref{fig:2-0-vortex-profiles}(d) is interpreted as a doubly quantized skyrmion~\cite{Yan2008.PRA77.033621,Mas2011.PRA84.033611} located at the circular interface of the two species, where $|\Psi_1|=|\Psi_2|$. Since $\windingnumber_\mathrm{s}=-2$, $\spinvector$ rotates by $-4\pi$ about $\unitzvector$ when the interface is traversed azimuthally; additionally, the projection $\unitzvector\cdot\spinvector$ changes from $-1$ to $1$ when the interface is crossed radially. For the $\tcv{2}{1}$ vortex in Fig.~\ref{fig:2-1-vortex-profiles}(d), the $\mathrm{U}\left(1\right)$ gauge winds by $3\pi$ and the spin $\spinvector$ by $-2\pi$. However, because now $\unitzvector\cdot\spinvector$ vanishes everywhere, this state is not a skyrmion but instead corresponds to a singly quantized \emph{spin vortex}. The defect is structurally similar to the so-called cross disgyration in the fermionic superfluid ${}^{3}$He-{\textit{A}}~\cite{Mak1977.JLTP27.635,Mer1979.RMP51.591}. Finally, the states in Fig.~\ref{fig:skyrmion-profiles} feature giant skyrmions with (a)~$\windingnumber_\mathrm{s}=-9$ and (b)~$\windingnumber_\mathrm{s}=-12$.

\section{Conclusion}

In this study, we have demonstrated that two-species BECs in rotating harmonic traps are able to host thermodynamically stable multiquantum (or giant) vortices. Such topological entities rarely exist in the ground state and have thus been elusive in BECs, whereas they are observed and useful in, e.g., mesoscopic superconductivity~\cite{Mel2002.Nat415.60,Sch1998.PRL81.2783,Kan2004.PRL93.257002,Gri2007.PRL99.147003,Cre2011.PRL107.097202}. In the present case, their stability is not induced by elaborate external potentials~\cite{Fis2003.PRL90.140402,Jos2004.Chaos14.875,Aft2004.PRA69.033608,Jac2004.PRA69.053619,Fet2005.PRA71.013605,Kim2005.PRA72.023619,Fu2006.PRA73.013614,Cor2007.JMP48.042104,Cor2011.PRA84.053614,Sim2002.PRA65.033614,Kuo2010.JLTP161.561,Kar2013.PRA87.043609} but is an inherent property of the harmonically confined, mass-imbalanced two-species system: The giant vortex in the heavier species is stabilized by its coupling to the lighter, giant-vortex-free species. 

Experimentally, the presence of the MQV could be verified, e.g., by measuring the orbital angular momentum using surface wave spectroscopy~\cite{Che2000.PRL85.2223,Hal2001.PRL86.2922,Lea2002.PRL89.190403} or by detecting the $\windingnumber_j$-dependent concentric density ripples that would form in free expansion~\cite{Seo2014.JKPS64.53}. Due to its ground-state nature, the MQV is expected to be highly reproducible, long lived, and therefore amenable to extensive measurements.

We also classified the discovered states into spin-skyrmion (coreless) and spin-vortex (cored) variants, both of which can be realized in a ${}^{87}$Rb--${}^{41}$K~BEC~\cite{Fer2002.PRL89.053202,Mod2002.PRL89.190404,Cat2008.PRA77.011603,Tha2008.PRL100.210402,Aik2009.NJP11.055035,Cat2009.PRL103.140401}. The similarities of these vortices with fractional~\cite{Geu2010.PRB81.214514} and skyrmionic~\cite{Gar2011.PRL107.197001} vortex states in multiband superconductors, as well as the rich possibilities for the creation and tuning of multispecies BECs~\cite{Fer2002.PRL89.053202,Mod2002.PRL89.190404,Cat2008.PRA77.011603,Tha2008.PRL100.210402,Aik2009.NJP11.055035,Cat2009.PRL103.140401,Car2011.PRA84.011603,Ler2011.EPJD65.3,Pas2013.PRA88.023601}, open a wide avenue for exploring emergent physics in multicomponent quantum systems consisting of inherently nonidentical components.

\begin{acknowledgments}
This work was supported by the Finnish Cultural Foundation, the Research Foundation - Flanders (FWO), and the Magnus Ehrnrooth Foundation. We thank E.~Ruokokoski and T.~P.~Simula for valuable comments and discussions.
\end{acknowledgments}

\bibliography{tc-gv-manu}

\begin{thebibliography}{10}%
\makeatletter
\providecommand \@ifxundefined [1]{%
 \ifx #1\undefined \expandafter \@firstoftwo
 \else \expandafter \@secondoftwo
\fi
}%
\providecommand \@ifnum [1]{%
 \ifnum #1\expandafter \@firstoftwo
 \else \expandafter \@secondoftwo
\fi
}%
\providecommand \enquote [1]{``#1''}%
\providecommand \bibnamefont  [1]{#1}%
\providecommand \bibfnamefont [1]{#1}%
\providecommand \citenamefont [1]{#1}%
\providecommand\href[0]{\@sanitize\@href}%
\providecommand\@href[1]{\endgroup\@@startlink{#1}\endgroup\@@href}%
\providecommand\@@href[1]{#1\@@endlink}%
\providecommand \@sanitize [0]{\begingroup\catcode`\&12\catcode`\#12\relax}%
\@ifxundefined \pdfoutput {\@firstoftwo}{%
 \@ifnum{\z@=\pdfoutput}{\@firstoftwo}{\@secondoftwo}%
}{%
 \providecommand\@@startlink[1]{\leavevmode\special{html:<a href="#1">}}%
 \providecommand\@@endlink[0]{\special{html:</a>}}%
}{%
 \providecommand\@@startlink[1]{%
  \leavevmode
  \pdfstartlink
   attr{/Border[0 0 1 ]/H/I/C[0 1 1]}%
   user{/Subtype/Link/A<</Type/Action/S/URI/URI(#1)>>}%
  \relax
 }%
 \providecommand\@@endlink[0]{\pdfendlink}%
}%
\providecommand \url  [0]{\begingroup\@sanitize \@url }%
\providecommand \@url [1]{\endgroup\@href {#1}{\urlprefix}}%
\providecommand \urlprefix [0]{URL }%
\providecommand \Eprint[0]{\href }%
\@ifxundefined \urlstyle {%
  \providecommand \doi [1]{doi:\discretionary{}{}{}#1}%
}{%
  \providecommand \doi [0]{doi:\discretionary{}{}{}\begingroup
  \urlstyle{rm}\Url }%
}%
\providecommand \doibase [0]{http://dx.doi.org/}%
\providecommand \Doi[1]{\href{\doibase#1}}%
\providecommand \bibAnnote [3]{%
  \BibitemShut{#1}%
  \begin{quotation}\noindent
    \textsc{Key:}\ #2\\\textsc{Annotation:}\ #3%
  \end{quotation}%
}%
\providecommand \bibAnnoteFile [2]{%
  \IfFileExists{#2}{\bibAnnote {#1} {#2} {\input{#2}}}{}%
}%
\providecommand \typeout [0]{\immediate \write \m@ne }%
\providecommand \selectlanguage [0]{\@gobble}%
\providecommand \bibinfo [0]{\@secondoftwo}%
\providecommand \bibfield [0]{\@secondoftwo}%
\providecommand \translation [1]{[#1]}%
\providecommand \BibitemOpen[0]{}%
\providecommand \bibitemStop [0]{}%
\providecommand \bibitemNoStop [0]{.\EOS\space}%
\providecommand \EOS [0]{\spacefactor3000\relax}%
\providecommand \BibitemShut [1]{\csname bibitem#1\endcsname}%
\bibitem{Ons1949.NCim6Sup2.249}%
  \BibitemOpen
  \bibfield{author}{%
  \bibinfo {author} {\bibfnamefont{L.}~\bibnamefont{Onsager}},\ }%
  \bibfield{journal}{%
  \bibinfo {journal} {Nuovo Cimento}\ }%
  \textbf{\bibinfo {volume} {6}},\ \bibinfo {pages} {Suppl. 2, 249} (\bibinfo
  {year} {1949}).%
  \bibAnnoteFile{NoStop}{Ons1949.NCim6Sup2.249}%
  \bibitem{Fey1955.PLTP1.17}%
  \BibitemOpen
  \bibfield{author}{%
  \bibinfo {author} {\bibfnamefont{R.~P.}\ \bibnamefont{Feynman}},\ }%
  \bibfield{journal}{%
  \Doi{10.1016/S0079-6417(08)60077-3}{\bibinfo {journal} {Prog. Low Temp. Phys.}}\ }%
  \textbf{\bibinfo {volume} {1}},\ \bibinfo {pages} {17} (\bibinfo {year}
  {1957}).%
  \bibAnnoteFile{NoStop}{Fey1955.PLTP1.17}%
\bibitem{Abr1957.ZETF32.1442}%
  \BibitemOpen
  \bibfield{author}{%
  \bibinfo {author} {\bibfnamefont{A.~A.}\ \bibnamefont{Abrikosov}},\ }%
  \bibfield{journal}{%
  \bibinfo {journal} {Zh. Eksp. Teor. Fiz.}\ }%
  \textbf{\bibinfo {volume} {32}},\ \bibinfo {pages} {1442} (\bibinfo {year}
  {1957})\ [\bibfield{journal}{%
  \bibinfo {journal} {Sov. Phys. JETP}\ }%
  \textbf{\bibinfo {volume} {5}},\ \bibinfo {pages} {1174} (\bibinfo {year}
  {1957})].%
  \bibAnnoteFile{NoStop}{Abr1957.ZETF32.1442}%
\bibitem{Yar1979.PRL43.214}%
  \BibitemOpen
  \bibfield{author}{%
  \bibinfo {author} {\bibfnamefont{E.~J.}\ \bibnamefont{Yarmchuk}}, \bibinfo
  {author} {\bibfnamefont{M.~J.~V.}\ \bibnamefont{Gordon}},\ and\ \bibinfo
  {author} {\bibfnamefont{R.~E.}\ \bibnamefont{Packard}},\ }%
  \bibfield{journal}{%
  \Doi{10.1103/PhysRevLett.43.214}{\bibinfo {journal} {Phys. Rev. Lett.}}\ }%
  \textbf{\bibinfo {volume} {43}},\ \bibinfo {pages} {214} (\bibinfo {year}
  {1979}).%
  \bibAnnoteFile{NoStop}{Yar1979.PRL43.214}%
\bibitem{Abo2001.Sci292.476}%
  \BibitemOpen
  \bibfield{author}{%
  \bibinfo {author} {\bibfnamefont{J.~R.}\ \bibnamefont{Abo-Shaeer}}, \bibinfo
  {author} {\bibfnamefont{C.}~\bibnamefont{Raman}}, \bibinfo {author}
  {\bibfnamefont{J.~M.}\ \bibnamefont{Vogels}},\ and\ \bibinfo {author}
  {\bibfnamefont{W.}~\bibnamefont{Ketterle}},\ }%
  \bibfield{journal}{%
  \Doi{10.1126/science.1060182}{\bibinfo {journal} {Science}}\ }%
  \textbf{\bibinfo {volume} {292}},\ \bibinfo {pages} {476} (\bibinfo {year}
  {2001}).%
  \bibAnnoteFile{NoStop}{Abo2001.Sci292.476}%
\bibitem{Zwi2005.Nat435.1047}%
  \BibitemOpen
  \bibfield{author}{%
  \bibinfo {author} {\bibfnamefont{M.~W.}\ \bibnamefont{Zwierlein}}, \bibinfo
  {author} {\bibfnamefont{J.~R.}\ \bibnamefont{Abo-Shaeer}}, \bibinfo {author}
  {\bibfnamefont{A.}~\bibnamefont{Schirotzek}}, \bibinfo {author}
  {\bibfnamefont{C.~H.}\ \bibnamefont{Schunck}},\ and\ \bibinfo {author}
  {\bibfnamefont{W.}~\bibnamefont{Ketterle}},\ }%
  \bibfield{journal}{%
  \Doi{10.1038/nature03858}{\bibinfo {journal} {Nature (London)}}\ }%
  \textbf{\bibinfo {volume} {435}},\ \bibinfo {pages} {1047} (\bibinfo {year}
  {2005}).%
  \bibAnnoteFile{NoStop}{Zwi2005.Nat435.1047}%
\bibitem{Kas2005.IJMPB19.1835}%
  \BibitemOpen
  \bibfield{author}{%
  \bibinfo {author} {\bibfnamefont{K.}~\bibnamefont{Kasamatsu}}, \bibinfo
  {author} {\bibfnamefont{M.}~\bibnamefont{Tsubota}},\ and\ \bibinfo {author}
  {\bibfnamefont{M.}~\bibnamefont{Ueda}},\ }%
  \bibfield{journal}{%
  \Doi{10.1142/S0217979205029602}{\bibinfo {journal} {Int. J. Mod. Phys. B}}\
  }%
  \textbf{\bibinfo {volume} {19}},\ \bibinfo {pages} {1835} (\bibinfo {year}
  {2005}).%
  \bibAnnoteFile{NoStop}{Kas2005.IJMPB19.1835}%
\bibitem{Mad2000.PRL84.806}%
  \BibitemOpen
  \bibfield{author}{%
  \bibinfo {author} {\bibfnamefont{K.~W.}\ \bibnamefont{Madison}}, \bibinfo
  {author} {\bibfnamefont{F.}~\bibnamefont{Chevy}}, \bibinfo {author}
  {\bibfnamefont{W.}~\bibnamefont{Wohlleben}},\ and\ \bibinfo {author}
  {\bibfnamefont{J.}~\bibnamefont{Dalibard}},\ }%
  \bibfield{journal}{%
  \Doi{10.1103/PhysRevLett.84.806}{\bibinfo {journal} {Phys. Rev. Lett.}}\ }%
  \textbf{\bibinfo {volume} {84}},\ \bibinfo {pages} {806} (\bibinfo {year}
  {2000}).%
  \bibAnnoteFile{NoStop}{Mad2000.PRL84.806}%
\bibitem{And2010.JLTP161.574}%
  \BibitemOpen
  \bibfield{author}{%
  \bibinfo {author} {\bibfnamefont{B.~P.}\ \bibnamefont{Anderson}},\ }%
  \bibfield{journal}{%
  \Doi{10.1007/s10909-010-0224-1}{\bibinfo {journal} {J. Low Temp. Phys.}}\ }%
  \textbf{\bibinfo {volume} {161}},\ \bibinfo {pages} {574} (\bibinfo {year}
  {2010}).%
  \bibAnnoteFile{NoStop}{And2010.JLTP161.574}%
\bibitem{Mat1999.PRL83.2498}%
  \BibitemOpen
  \bibfield{author}{%
  \bibinfo {author} {\bibfnamefont{M.~R.}\ \bibnamefont{Matthews}}, \bibinfo
  {author} {\bibfnamefont{B.~P.}\ \bibnamefont{Anderson}}, \bibinfo {author}
  {\bibfnamefont{P.~C.}\ \bibnamefont{Haljan}}, \bibinfo {author}
  {\bibfnamefont{D.~S.}\ \bibnamefont{Hall}}, \bibinfo {author}
  {\bibfnamefont{C.~E.}\ \bibnamefont{Wieman}},\ and\ \bibinfo {author}
  {\bibfnamefont{E.~A.}\ \bibnamefont{Cornell}},\ }%
  \bibfield{journal}{%
  \Doi{10.1103/PhysRevLett.83.2498}{\bibinfo {journal} {Phys. Rev. Lett.}}\ }%
  \textbf{\bibinfo {volume} {83}},\ \bibinfo {pages} {2498} (\bibinfo {year}
  {1999}).%
  \bibAnnoteFile{NoStop}{Mat1999.PRL83.2498}%
\bibitem{Sch2004.PRL93.210403}%
  \BibitemOpen
  \bibfield{author}{%
  \bibinfo {author} {\bibfnamefont{V.}~\bibnamefont{Schweikhard}}, \bibinfo
  {author} {\bibfnamefont{I.}~\bibnamefont{Coddington}}, \bibinfo {author}
  {\bibfnamefont{P.}~\bibnamefont{Engels}}, \bibinfo {author}
  {\bibfnamefont{S.}~\bibnamefont{Tung}},\ and\ \bibinfo {author}
  {\bibfnamefont{E.~A.}\ \bibnamefont{Cornell}},\ }%
  \bibfield{journal}{%
  \Doi{10.1103/PhysRevLett.93.210403}{\bibinfo {journal} {Phys. Rev. Lett.}}\
  }%
  \textbf{\bibinfo {volume} {93}},\ \bibinfo {pages} {210403} (\bibinfo {year}
  {2004}).%
  \bibAnnoteFile{NoStop}{Sch2004.PRL93.210403}%
\bibitem{Mya1997.PRL78.586}%
  \BibitemOpen
  \bibfield{author}{%
  \bibinfo {author} {\bibfnamefont{C.~J.}\ \bibnamefont{Myatt}}, \bibinfo
  {author} {\bibfnamefont{E.~A.}\ \bibnamefont{Burt}}, \bibinfo {author}
  {\bibfnamefont{R.~W.}\ \bibnamefont{Ghrist}}, \bibinfo {author}
  {\bibfnamefont{E.~A.}\ \bibnamefont{Cornell}},\ and\ \bibinfo {author}
  {\bibfnamefont{C.~E.}\ \bibnamefont{Wieman}},\ }%
  \bibfield{journal}{%
  \Doi{10.1103/PhysRevLett.78.586}{\bibinfo {journal} {Phys. Rev. Lett.}}\ }%
  \textbf{\bibinfo {volume} {78}},\ \bibinfo {pages} {586} (\bibinfo {year}
  {1997}).%
  \bibAnnoteFile{NoStop}{Mya1997.PRL78.586}%
\bibitem{Hal1998.PRL81.1539}%
  \BibitemOpen
  \bibfield{author}{%
  \bibinfo {author} {\bibfnamefont{D.~S.}\ \bibnamefont{Hall}}, \bibinfo
  {author} {\bibfnamefont{M.~R.}\ \bibnamefont{Matthews}}, \bibinfo {author}
  {\bibfnamefont{J.~R.}\ \bibnamefont{Ensher}}, \bibinfo {author}
  {\bibfnamefont{C.~E.}\ \bibnamefont{Wieman}},\ and\ \bibinfo {author}
  {\bibfnamefont{E.~A.}\ \bibnamefont{Cornell}},\ }%
  \bibfield{journal}{%
  \Doi{10.1103/PhysRevLett.81.1539}{\bibinfo {journal} {Phys. Rev. Lett.}}\ }%
  \textbf{\bibinfo {volume} {81}},\ \bibinfo {pages} {1539} (\bibinfo {year}
  {1998}).%
  \bibAnnoteFile{NoStop}{Hal1998.PRL81.1539}%
\bibitem{Del2001.PRA63.051602}%
  \BibitemOpen
  \bibfield{author}{%
  \bibinfo {author} {\bibfnamefont{G.}~\bibnamefont{Delannoy}}, \bibinfo
  {author} {\bibfnamefont{S.~G.}\ \bibnamefont{Murdoch}}, \bibinfo {author}
  {\bibfnamefont{V.}~\bibnamefont{Boyer}}, \bibinfo {author}
  {\bibfnamefont{V.}~\bibnamefont{Josse}}, \bibinfo {author}
  {\bibfnamefont{P.}~\bibnamefont{Bouyer}},\ and\ \bibinfo {author}
  {\bibfnamefont{A.}~\bibnamefont{Aspect}},\ }%
  \bibfield{journal}{%
  \Doi{10.1103/PhysRevA.63.051602}{\bibinfo {journal} {Phys. Rev. A}}\ }%
  \textbf{\bibinfo {volume} {63}},\ \bibinfo {pages} {051602} (\bibinfo {year}
  {2001}).%
  \bibAnnoteFile{NoStop}{Del2001.PRA63.051602}%
\bibitem{And2009.PRA80.023603}%
  \BibitemOpen
  \bibfield{author}{%
  \bibinfo {author} {\bibfnamefont{R.~P.}\ \bibnamefont{Anderson}}, \bibinfo
  {author} {\bibfnamefont{C.}~\bibnamefont{Ticknor}}, \bibinfo {author}
  {\bibfnamefont{A.~I.}\ \bibnamefont{Sidorov}},\ and\ \bibinfo {author}
  {\bibfnamefont{B.~V.}\ \bibnamefont{Hall}},\ }%
  \bibfield{journal}{%
  \Doi{10.1103/PhysRevA.80.023603}{\bibinfo {journal} {Phys. Rev. A}}\ }%
  \textbf{\bibinfo {volume} {80}},\ \bibinfo {pages} {023603} (\bibinfo {year}
  {2009}).%
  \bibAnnoteFile{NoStop}{And2009.PRA80.023603}%
\bibitem{Bap2008.PRL101.040402}%
  \BibitemOpen
  \bibfield{author}{%
  \bibinfo {author} {\bibfnamefont{S.~B.}\ \bibnamefont{Papp}}, \bibinfo
  {author} {\bibfnamefont{J.~M.}\ \bibnamefont{Pino}},\ and\ \bibinfo {author}
  {\bibfnamefont{C.~E.}\ \bibnamefont{Wieman}},\ }%
  \bibfield{journal}{%
  \Doi{10.1103/PhysRevLett.101.040402}{\bibinfo {journal} {Phys. Rev. Lett.}}\
  }%
  \textbf{\bibinfo {volume} {101}},\ \bibinfo {pages} {040402} (\bibinfo {year}
  {2008}).%
  \bibAnnoteFile{NoStop}{Bap2008.PRL101.040402}%
\bibitem{Sug2011.PRA84.011610}%
  \BibitemOpen
  \bibfield{author}{%
  \bibinfo {author} {\bibfnamefont{S.}~\bibnamefont{Sugawa}}, \bibinfo {author}
  {\bibfnamefont{R.}~\bibnamefont{Yamazaki}}, \bibinfo {author}
  {\bibfnamefont{S.}~\bibnamefont{Taie}},\ and\ \bibinfo {author}
  {\bibfnamefont{Y.}~\bibnamefont{Takahashi}},\ }%
  \bibfield{journal}{%
  \Doi{10.1103/PhysRevA.84.011610}{\bibinfo {journal} {Phys. Rev. A}}\ }%
  \textbf{\bibinfo {volume} {84}},\ \bibinfo {pages} {011610} (\bibinfo {year}
  {2011}).%
  \bibAnnoteFile{NoStop}{Sug2011.PRA84.011610}%
\bibitem{Fer2002.PRL89.053202}%
  \BibitemOpen
  \bibfield{author}{%
  \bibinfo {author} {\bibfnamefont{G.}~\bibnamefont{Ferrari}}, \bibinfo
  {author} {\bibfnamefont{M.}~\bibnamefont{Inguscio}}, \bibinfo {author}
  {\bibfnamefont{W.}~\bibnamefont{Jastrzebski}}, \bibinfo {author}
  {\bibfnamefont{G.}~\bibnamefont{Modugno}}, \bibinfo {author}
  {\bibfnamefont{G.}~\bibnamefont{Roati}},\ and\ \bibinfo {author}
  {\bibfnamefont{A.}~\bibnamefont{Simoni}},\ }%
  \bibfield{journal}{%
  \Doi{10.1103/PhysRevLett.89.053202}{\bibinfo {journal} {Phys. Rev. Lett.}}\
  }%
  \textbf{\bibinfo {volume} {89}},\ \bibinfo {pages} {053202} (\bibinfo {year}
  {2002}).%
  \bibAnnoteFile{NoStop}{Fer2002.PRL89.053202}%
\bibitem{Mod2002.PRL89.190404}%
  \BibitemOpen
  \bibfield{author}{%
  \bibinfo {author} {\bibfnamefont{G.}~\bibnamefont{Modugno}}, \bibinfo
  {author} {\bibfnamefont{M.}~\bibnamefont{Modugno}}, \bibinfo {author}
  {\bibfnamefont{F.}~\bibnamefont{Riboli}}, \bibinfo {author}
  {\bibfnamefont{G.}~\bibnamefont{Roati}},\ and\ \bibinfo {author}
  {\bibfnamefont{M.}~\bibnamefont{Inguscio}},\ }%
  \bibfield{journal}{%
  \Doi{10.1103/PhysRevLett.89.190404}{\bibinfo {journal} {Phys. Rev. Lett.}}\
  }%
  \textbf{\bibinfo {volume} {89}},\ \bibinfo {pages} {190404} (\bibinfo {year}
  {2002}).%
  \bibAnnoteFile{NoStop}{Mod2002.PRL89.190404}%
\bibitem{Cat2008.PRA77.011603}%
  \BibitemOpen
  \bibfield{author}{%
  \bibinfo {author} {\bibfnamefont{J.}~\bibnamefont{Catani}}, \bibinfo {author}
  {\bibfnamefont{L.}~\bibnamefont{De~Sarlo}}, \bibinfo {author}
  {\bibfnamefont{G.}~\bibnamefont{Barontini}}, \bibinfo {author}
  {\bibfnamefont{F.}~\bibnamefont{Minardi}},\ and\ \bibinfo {author}
  {\bibfnamefont{M.}~\bibnamefont{Inguscio}},\ }%
  \bibfield{journal}{%
  \Doi{10.1103/PhysRevA.77.011603}{\bibinfo {journal} {Phys. Rev. A}}\ }%
  \textbf{\bibinfo {volume} {77}},\ \bibinfo {pages} {011603} (\bibinfo {year}
  {2008}).%
  \bibAnnoteFile{NoStop}{Cat2008.PRA77.011603}%
\bibitem{Tha2008.PRL100.210402}%
  \BibitemOpen
  \bibfield{author}{%
  \bibinfo {author} {\bibfnamefont{G.}~\bibnamefont{Thalhammer}}, \bibinfo
  {author} {\bibfnamefont{G.}~\bibnamefont{Barontini}}, \bibinfo {author}
  {\bibfnamefont{L.}~\bibnamefont{De~Sarlo}}, \bibinfo {author}
  {\bibfnamefont{J.}~\bibnamefont{Catani}}, \bibinfo {author}
  {\bibfnamefont{F.}~\bibnamefont{Minardi}},\ and\ \bibinfo {author}
  {\bibfnamefont{M.}~\bibnamefont{Inguscio}},\ }%
  \bibfield{journal}{%
  \Doi{10.1103/PhysRevLett.100.210402}{\bibinfo {journal} {Phys. Rev. Lett.}}\
  }%
  \textbf{\bibinfo {volume} {100}},\ \bibinfo {pages} {210402} (\bibinfo {year}
  {2008}).%
  \bibAnnoteFile{NoStop}{Tha2008.PRL100.210402}%
\bibitem{Aik2009.NJP11.055035}%
  \BibitemOpen
  \bibfield{author}{%
  \bibinfo {author} {\bibfnamefont{K.}~\bibnamefont{Aikawa}}, \bibinfo {author}
  {\bibfnamefont{D.}~\bibnamefont{Akamatsu}}, \bibinfo {author}
  {\bibfnamefont{J.}~\bibnamefont{Kobayashi}}, \bibinfo {author}
  {\bibfnamefont{M.}~\bibnamefont{Ueda}}, \bibinfo {author}
  {\bibfnamefont{T.}~\bibnamefont{Kishimoto}},\ and\ \bibinfo {author}
  {\bibfnamefont{S.}~\bibnamefont{Inouye}},\ }%
  \bibfield{journal}{%
  \Doi{10.1088/1367-2630/11/5/055035}{\bibinfo {journal} {New J. Phys.}}\ }%
  \textbf{\bibinfo {volume} {11}},\ \bibinfo {pages} {055035} (\bibinfo {year}
  {2009}).%
  \bibAnnoteFile{NoStop}{Aik2009.NJP11.055035}%
\bibitem{Cat2009.PRL103.140401}%
  \BibitemOpen
  \bibfield{author}{%
  \bibinfo {author} {\bibfnamefont{J.}~\bibnamefont{Catani}}, \bibinfo {author}
  {\bibfnamefont{G.}~\bibnamefont{Barontini}}, \bibinfo {author}
  {\bibfnamefont{G.}~\bibnamefont{Lamporesi}}, \bibinfo {author}
  {\bibfnamefont{F.}~\bibnamefont{Rabatti}}, \bibinfo {author}
  {\bibfnamefont{G.}~\bibnamefont{Thalhammer}}, \bibinfo {author}
  {\bibfnamefont{F.}~\bibnamefont{Minardi}}, \bibinfo {author}
  {\bibfnamefont{S.}~\bibnamefont{Stringari}},\ and\ \bibinfo {author}
  {\bibfnamefont{M.}~\bibnamefont{Inguscio}},\ }%
  \bibfield{journal}{%
  \Doi{10.1103/PhysRevLett.103.140401}{\bibinfo {journal} {Phys. Rev. Lett.}}\
  }%
  \textbf{\bibinfo {volume} {103}},\ \bibinfo {pages} {140401} (\bibinfo {year}
  {2009}).%
  \bibAnnoteFile{NoStop}{Cat2009.PRL103.140401}%
\bibitem{Car2011.PRA84.011603}%
  \BibitemOpen
  \bibfield{author}{%
  \bibinfo {author} {\bibfnamefont{D.~J.}\ \bibnamefont{McCarron}}, \bibinfo
  {author} {\bibfnamefont{H.~W.}\ \bibnamefont{Cho}}, \bibinfo {author}
  {\bibfnamefont{D.~L.}\ \bibnamefont{Jenkin}}, \bibinfo {author}
  {\bibfnamefont{M.~P.}\ \bibnamefont{K\"oppinger}},\ and\ \bibinfo {author}
  {\bibfnamefont{S.~L.}\ \bibnamefont{Cornish}},\ }%
  \bibfield{journal}{%
  \Doi{10.1103/PhysRevA.84.011603}{\bibinfo {journal} {Phys. Rev. A}}\ }%
  \textbf{\bibinfo {volume} {84}},\ \bibinfo {pages} {011603} (\bibinfo {year}
  {2011}).%
  \bibAnnoteFile{NoStop}{Car2011.PRA84.011603}%
\bibitem{Ler2011.EPJD65.3}%
  \BibitemOpen
  \bibfield{author}{%
  \bibinfo {author} {\bibfnamefont{A.}~\bibnamefont{Lercher}}, \bibinfo
  {author} {\bibfnamefont{T.}~\bibnamefont{Takekoshi}}, \bibinfo {author}
  {\bibfnamefont{M.}~\bibnamefont{Debatin}}, \bibinfo {author}
  {\bibfnamefont{B.}~\bibnamefont{Schuster}}, \bibinfo {author}
  {\bibfnamefont{R.}~\bibnamefont{Rameshan}}, \bibinfo {author}
  {\bibfnamefont{F.}~\bibnamefont{Ferlaino}}, \bibinfo {author}
  {\bibfnamefont{R.}~\bibnamefont{Grimm}},\ and\ \bibinfo {author}
  {\bibfnamefont{H.-C.}\ \bibnamefont{N\"agerl}},\ }%
  \bibfield{journal}{%
  \Doi{10.1140/epjd/e2011-20015-6}{\bibinfo {journal} {Eur. Phys. J. D}}\ }%
  \textbf{\bibinfo {volume} {65}},\ \bibinfo {pages} {3} (\bibinfo {year}
  {2011}).%
  \bibAnnoteFile{NoStop}{Ler2011.EPJD65.3}%
\bibitem{Pas2013.PRA88.023601}%
  \BibitemOpen
  \bibfield{author}{%
  \bibinfo {author} {\bibfnamefont{B.}~\bibnamefont{Pasquiou}}, \bibinfo
  {author} {\bibfnamefont{A.}~\bibnamefont{Bayerle}}, \bibinfo {author}
  {\bibfnamefont{S.~M.}\ \bibnamefont{Tzanova}}, \bibinfo {author}
  {\bibfnamefont{S.}~\bibnamefont{Stellmer}}, \bibinfo {author}
  {\bibfnamefont{J.}~\bibnamefont{Szczepkowski}}, \bibinfo {author}
  {\bibfnamefont{M.}~\bibnamefont{Parigger}}, \bibinfo {author}
  {\bibfnamefont{R.}~\bibnamefont{Grimm}},\ and\ \bibinfo {author}
  {\bibfnamefont{F.}~\bibnamefont{Schreck}},\ }%
  \bibfield{journal}{%
  \Doi{10.1103/PhysRevA.88.023601}{\bibinfo {journal} {Phys. Rev. A}}\ }%
  \textbf{\bibinfo {volume} {88}},\ \bibinfo {pages} {023601} (\bibinfo {year}
  {2013}).%
  \bibAnnoteFile{NoStop}{Pas2013.PRA88.023601}%
\bibitem{Kas2004.PRL93.250406}%
  \BibitemOpen
  \bibfield{author}{%
  \bibinfo {author} {\bibfnamefont{K.}~\bibnamefont{Kasamatsu}}, \bibinfo
  {author} {\bibfnamefont{M.}~\bibnamefont{Tsubota}},\ and\ \bibinfo {author}
  {\bibfnamefont{M.}~\bibnamefont{Ueda}},\ }%
  \bibfield{journal}{%
  \Doi{10.1103/PhysRevLett.93.250406}{\bibinfo {journal} {Phys. Rev. Lett.}}\
  }%
  \textbf{\bibinfo {volume} {93}},\ \bibinfo {pages} {250406} (\bibinfo {year}
  {2004}).%
  \bibAnnoteFile{NoStop}{Kas2004.PRL93.250406}%
\bibitem{Kas2005.PRA71.043611}%
  \BibitemOpen
  \bibfield{author}{%
  \bibinfo {author} {\bibfnamefont{K.}~\bibnamefont{Kasamatsu}}, \bibinfo
  {author} {\bibfnamefont{M.}~\bibnamefont{Tsubota}},\ and\ \bibinfo {author}
  {\bibfnamefont{M.}~\bibnamefont{Ueda}},\ }%
  \bibfield{journal}{%
  \Doi{10.1103/PhysRevA.71.043611}{\bibinfo {journal} {Phys. Rev. A}}\ }%
  \textbf{\bibinfo {volume} {71}},\ \bibinfo {pages} {043611} (\bibinfo {year}
  {2005}).%
  \bibAnnoteFile{NoStop}{Kas2005.PRA71.043611}%
\bibitem{Mas2011.PRA84.033611}%
  \BibitemOpen
  \bibfield{author}{%
  \bibinfo {author} {\bibfnamefont{P.}~\bibnamefont{Mason}}\ and\ \bibinfo
  {author} {\bibfnamefont{A.}~\bibnamefont{Aftalion}},\ }%
  \bibfield{journal}{%
  \Doi{10.1103/PhysRevA.84.033611}{\bibinfo {journal} {Phys. Rev. A}}\ }%
  \textbf{\bibinfo {volume} {84}},\ \bibinfo {pages} {033611} (\bibinfo {year}
  {2011}).%
  \bibAnnoteFile{NoStop}{Mas2011.PRA84.033611}%
\bibitem{Mue2002.PRL88.180403}%
  \BibitemOpen
  \bibfield{author}{%
  \bibinfo {author} {\bibfnamefont{E.~J.}\ \bibnamefont{Mueller}}\ and\
  \bibinfo {author} {\bibfnamefont{T.-L.}\ \bibnamefont{Ho}},\ }%
  \bibfield{journal}{%
  \Doi{10.1103/PhysRevLett.88.180403}{\bibinfo {journal} {Phys. Rev. Lett.}}\
  }%
  \textbf{\bibinfo {volume} {88}},\ \bibinfo {pages} {180403} (\bibinfo {year}
  {2002}).%
  \bibAnnoteFile{NoStop}{Mue2002.PRL88.180403}%
\bibitem{Kas2003.PRL91.150406}%
  \BibitemOpen
  \bibfield{author}{%
  \bibinfo {author} {\bibfnamefont{K.}~\bibnamefont{Kasamatsu}}, \bibinfo
  {author} {\bibfnamefont{M.}~\bibnamefont{Tsubota}},\ and\ \bibinfo {author}
  {\bibfnamefont{M.}~\bibnamefont{Ueda}},\ }%
  \bibfield{journal}{%
  \Doi{10.1103/PhysRevLett.91.150406}{\bibinfo {journal} {Phys. Rev. Lett.}}\
  }%
  \textbf{\bibinfo {volume} {91}},\ \bibinfo {pages} {150406} (\bibinfo {year}
  {2003}).%
  \bibAnnoteFile{NoStop}{Kas2003.PRL91.150406}%
\bibitem{Kec2006.PRA73.023611}%
  \BibitemOpen
  \bibfield{author}{%
  \bibinfo {author} {\bibfnamefont{M.}~\bibnamefont{Ke{\ifmmode
  \mbox{\c{c}}\else \c{c}\fi{}}eli}}\ and\ \bibinfo {author}
  {\bibfnamefont{M.~\"O.}\ \bibnamefont{Oktel}},\ }%
  \bibfield{journal}{%
  \Doi{10.1103/PhysRevA.73.023611}{\bibinfo {journal} {Phys. Rev. A}}\ }%
  \textbf{\bibinfo {volume} {73}},\ \bibinfo {pages} {023611} (\bibinfo {year}
  {2006}).%
  \bibAnnoteFile{NoStop}{Kec2006.PRA73.023611}%
\bibitem{Min2009.PRA79.013605}%
  \BibitemOpen
  \bibfield{author}{%
  \bibinfo {author} {\bibfnamefont{M.~P.}\ \bibnamefont{Mink}}, \bibinfo
  {author} {\bibfnamefont{C.~M.}\ \bibnamefont{Smith}},\ and\ \bibinfo {author}
  {\bibfnamefont{R.~A.}\ \bibnamefont{Duine}},\ }%
  \bibfield{journal}{%
  \Doi{10.1103/PhysRevA.79.013605}{\bibinfo {journal} {Phys. Rev. A}}\ }%
  \textbf{\bibinfo {volume} {79}},\ \bibinfo {pages} {013605} (\bibinfo {year}
  {2009}).%
  \bibAnnoteFile{NoStop}{Min2009.PRA79.013605}%
\bibitem{Dah2008.PRB78.144510}%
  \BibitemOpen
  \bibfield{author}{%
  \bibinfo {author} {\bibfnamefont{E.~K.}\ \bibnamefont{Dahl}}, \bibinfo
  {author} {\bibfnamefont{E.}~\bibnamefont{Babaev}},\ and\ \bibinfo {author}
  {\bibfnamefont{A.}~\bibnamefont{Sudb\o{}}},\ }%
  \bibfield{journal}{%
  \Doi{10.1103/PhysRevB.78.144510}{\bibinfo {journal} {Phys. Rev. B}}\ }%
  \textbf{\bibinfo {volume} {78}},\ \bibinfo {pages} {144510} (\bibinfo {year}
  {2008}).%
  \bibAnnoteFile{NoStop}{Dah2008.PRB78.144510}%
\bibitem{Kuo2012.PRA85.043613}%
  \BibitemOpen
  \bibfield{author}{%
  \bibinfo {author} {\bibfnamefont{P.}~\bibnamefont{Kuopanportti}}, \bibinfo
  {author} {\bibfnamefont{J.~A.~M.}\ \bibnamefont{Huhtam\"aki}},\ and\ \bibinfo
  {author} {\bibfnamefont{M.}~\bibnamefont{M\"ott\"onen}},\ }%
  \bibfield{journal}{%
  \Doi{10.1103/PhysRevA.85.043613}{\bibinfo {journal} {Phys. Rev. A}}\ }%
  \textbf{\bibinfo {volume} {85}},\ \bibinfo {pages} {043613} (\bibinfo {year}
  {2012}).%
  \bibAnnoteFile{NoStop}{Kuo2012.PRA85.043613}%
\bibitem{Kas2009.PRA79.023606}%
  \BibitemOpen
  \bibfield{author}{%
  \bibinfo {author} {\bibfnamefont{K.}~\bibnamefont{Kasamatsu}}\ and\ \bibinfo
  {author} {\bibfnamefont{M.}~\bibnamefont{Tsubota}},\ }%
  \bibfield{journal}{%
  \Doi{10.1103/PhysRevA.79.023606}{\bibinfo {journal} {Phys. Rev. A}}\ }%
  \textbf{\bibinfo {volume} {79}},\ \bibinfo {pages} {023606} (\bibinfo {year}
  {2009}).%
  \bibAnnoteFile{NoStop}{Kas2009.PRA79.023606}%
\bibitem{Yan2008.PRA77.033621}%
  \BibitemOpen
  \bibfield{author}{%
  \bibinfo {author} {\bibfnamefont{S.-J.}\ \bibnamefont{Yang}}, \bibinfo
  {author} {\bibfnamefont{Q.-S.}\ \bibnamefont{Wu}}, \bibinfo {author}
  {\bibfnamefont{S.-N.}\ \bibnamefont{Zhang}},\ and\ \bibinfo {author}
  {\bibfnamefont{S.}~\bibnamefont{Feng}},\ }%
  \bibfield{journal}{%
  \Doi{10.1103/PhysRevA.77.033621}{\bibinfo {journal} {Phys. Rev. A}}\ }%
  \textbf{\bibinfo {volume} {77}},\ \bibinfo {pages} {033621} (\bibinfo {year}
  {2008}).%
  \bibAnnoteFile{NoStop}{Yan2008.PRA77.033621}%
\bibitem{Ron2011.SciRep1.43}%
  \BibitemOpen
  \bibfield{author}{%
  \bibinfo {author} {\bibfnamefont{M.}~\bibnamefont{Roncaglia}}, \bibinfo
  {author} {\bibfnamefont{M.}~\bibnamefont{Rizzi}},\ and\ \bibinfo {author}
  {\bibfnamefont{J.}~\bibnamefont{Dalibard}},\ }%
  \bibfield{journal}{%
  \Doi{10.1038/srep00043}{\bibinfo {journal} {Sci. Rep.}}\ }%
  \textbf{\bibinfo {volume} {1}},\ \bibinfo {pages} {43} (\bibinfo {year}
  {2011}).%
  \bibAnnoteFile{NoStop}{Ron2011.SciRep1.43}%
\bibitem{Abr1995.PRB52.7018}%
  \BibitemOpen
  \bibfield{author}{%
  \bibinfo {author} {\bibfnamefont{M.}~\bibnamefont{Abraham}}, \bibinfo
  {author} {\bibfnamefont{I.}~\bibnamefont{Aranson}},\ and\ \bibinfo {author}
  {\bibfnamefont{B.}~\bibnamefont{Galanti}},\ }%
  \bibfield{journal}{%
  \Doi{10.1103/PhysRevB.52.R7018}{\bibinfo {journal} {Phys. Rev. B}}\ }%
  \textbf{\bibinfo {volume} {52}},\ \bibinfo {pages} {R7018} (\bibinfo {year}
  {1995}).%
  \bibAnnoteFile{NoStop}{Abr1995.PRB52.7018}%
\bibitem{Ara1996.PRB53.75}%
  \BibitemOpen
  \bibfield{author}{%
  \bibinfo {author} {\bibfnamefont{I.}~\bibnamefont{Aranson}}\ and\ \bibinfo
  {author} {\bibfnamefont{V.}~\bibnamefont{Steinberg}},\ }%
  \bibfield{journal}{%
  \Doi{10.1103/PhysRevB.53.75}{\bibinfo {journal} {Phys. Rev. B}}\ }%
  \textbf{\bibinfo {volume} {53}},\ \bibinfo {pages} {75} (\bibinfo {year}
  {1996}).%
  \bibAnnoteFile{NoStop}{Ara1996.PRB53.75}%
\bibitem{San2014.arXiv.1405.0992}%
  \BibitemOpen
  \bibfield{author}{%
  \bibinfo {author} {\bibfnamefont{A.~C.}\ \bibnamefont{Santos}}, \bibinfo
  {author} {\bibfnamefont{V.~S.}\ \bibnamefont{Bagnato}},\ and\ \bibinfo
  {author} {\bibfnamefont{F.~E.~A.}\ \bibnamefont{dos Santos}},\ }%
  \bibinfo {note} {e-print {\href{http://arxiv.org/abs/1405.0992}{arXiv:1405.0992}}}.%
  \bibAnnoteFile{NoStop}{San2014.arXiv.1405.0992}%
\bibitem{Mel2002.Nat415.60}%
  \BibitemOpen
  \bibfield{author}{%
  \bibinfo {author} {\bibfnamefont{A.~S.}~\bibnamefont{Mel'nikov}}\ and\ \bibinfo
  {author} {\bibfnamefont{V.~M.}~\bibnamefont{Vinokur}},\ }%
  \bibfield{journal}{%
  \Doi{10.1038/415060a}{\bibinfo {journal} {Nature (London)}}\ }%
  \textbf{\bibinfo {volume} {415}},\ \bibinfo {pages} {60} (\bibinfo {year}
  {2002}).%
  \bibAnnoteFile{NoStop}{Mel2002.Nat415.60}%
\bibitem{Sch1998.PRL81.2783}%
  \BibitemOpen
  \bibfield{author}{%
  \bibinfo {author} {\bibfnamefont{V.~A.}\ \bibnamefont{Schweigert}}, \bibinfo
  {author} {\bibfnamefont{F.~M.}\ \bibnamefont{Peeters}},\ and\ \bibinfo
  {author} {\bibfnamefont{P.~S.}\ \bibnamefont{Deo}},\ }%
  \bibfield{journal}{%
  \Doi{10.1103/PhysRevLett.81.2783}{\bibinfo {journal} {Phys. Rev. Lett.}}\ }%
  \textbf{\bibinfo {volume} {81}},\ \bibinfo {pages} {2783} (\bibinfo {year}
  {1998}).%
  \bibAnnoteFile{NoStop}{Sch1998.PRL81.2783}%
\bibitem{Kan2004.PRL93.257002}%
  \BibitemOpen
  \bibfield{author}{%
  \bibinfo {author} {\bibfnamefont{A.}~\bibnamefont{Kanda}}, \bibinfo {author}
  {\bibfnamefont{B.~J.}\ \bibnamefont{Baelus}}, \bibinfo {author}
  {\bibfnamefont{F.~M.}\ \bibnamefont{Peeters}}, \bibinfo {author}
  {\bibfnamefont{K.}~\bibnamefont{Kadowaki}},\ and\ \bibinfo {author}
  {\bibfnamefont{Y.}~\bibnamefont{Ootuka}},\ }%
  \bibfield{journal}{%
  \Doi{10.1103/PhysRevLett.93.257002}{\bibinfo {journal} {Phys. Rev. Lett.}}\
  }%
  \textbf{\bibinfo {volume} {93}},\ \bibinfo {pages} {257002} (\bibinfo {year}
  {2004}).%
  \bibAnnoteFile{NoStop}{Kan2004.PRL93.257002}%
\bibitem{Gri2007.PRL99.147003}%
  \BibitemOpen
  \bibfield{author}{%
  \bibinfo {author} {\bibfnamefont{I.~V.}\ \bibnamefont{Grigorieva}}, \bibinfo
  {author} {\bibfnamefont{W.}~\bibnamefont{Escoffier}}, \bibinfo {author}
  {\bibfnamefont{V.~R.}\ \bibnamefont{Misko}}, \bibinfo {author}
  {\bibfnamefont{B.~J.}\ \bibnamefont{Baelus}}, \bibinfo {author}
  {\bibfnamefont{F.~M.}\ \bibnamefont{Peeters}}, \bibinfo {author}
  {\bibfnamefont{L.~Y.}\ \bibnamefont{Vinnikov}},\ and\ \bibinfo {author}
  {\bibfnamefont{S.~V.}\ \bibnamefont{Dubonos}},\ }%
  \bibfield{journal}{%
  \Doi{10.1103/PhysRevLett.99.147003}{\bibinfo {journal} {Phys. Rev. Lett.}}\
  }%
  \textbf{\bibinfo {volume} {99}},\ \bibinfo {pages} {147003} (\bibinfo {year}
  {2007}).%
  \bibAnnoteFile{NoStop}{Gri2007.PRL99.147003}%
\bibitem{Cre2011.PRL107.097202}%
  \BibitemOpen
  \bibfield{author}{%
  \bibinfo {author} {\bibfnamefont{T.}~\bibnamefont{Cren}}, \bibinfo {author}
  {\bibfnamefont{L.}~\bibnamefont{Serrier-Garcia}}, \bibinfo {author}
  {\bibfnamefont{F.}~\bibnamefont{Debontridder}},\ and\ \bibinfo {author}
  {\bibfnamefont{D.}~\bibnamefont{Roditchev}},\ }%
  \bibfield{journal}{%
  \Doi{10.1103/PhysRevLett.107.097202}{\bibinfo {journal} {Phys. Rev. Lett.}}\
  }%
  \textbf{\bibinfo {volume} {107}},\ \bibinfo {pages} {097202} (\bibinfo {year}
  {2011}).%
  \bibAnnoteFile{NoStop}{Cre2011.PRL107.097202}%
  \bibitem{note_circulation}%
  \BibitemOpen
  \bibinfo {note} {Essentially, this is because the circulation $\oint \mathbf{v}\left(\mathbf{r} \right) \cdot \,\mathrm{d}\mathbf{r}$ of the superfluid velocity $\mathbf{v}=\hbar\nabla\mathrm{Arg}\left(\Psi\right)/m$ is quantized in units of $2\pi\hbar/m$, where $m$ is the mass of the superfluid particles.}%
  \bibAnnoteFile{Stop}{note_circulation}%
\bibitem{Geu2008.PRA78.053610}%
  \BibitemOpen
  \bibfield{author}{%
  \bibinfo {author} {\bibfnamefont{R.}~\bibnamefont{Geurts}}, \bibinfo {author}
  {\bibfnamefont{M.~V.}\ \bibnamefont{Milo\ifmmode \check{s}\else
  \v{s}\fi{}evi\ifmmode~\acute{c}\else \'{c}\fi{}}},\ and\ \bibinfo {author}
  {\bibfnamefont{F.~M.}\ \bibnamefont{Peeters}},\ }%
  \bibfield{journal}{%
  \Doi{10.1103/PhysRevA.78.053610}{\bibinfo {journal} {Phys. Rev. A}}\ }%
  \textbf{\bibinfo {volume} {78}},\ \bibinfo {pages} {053610} (\bibinfo {year}
  {2008}).%
  \bibAnnoteFile{NoStop}{Geu2008.PRA78.053610}%
\bibitem{Ho1978.PRB18.1144}%
  \BibitemOpen
  \bibfield{author}{%
  \bibinfo {author} {\bibfnamefont{T.-L.}\ \bibnamefont{Ho}},\ }%
  \bibfield{journal}{%
  \Doi{10.1103/PhysRevB.18.1144}{\bibinfo {journal} {Phys. Rev. B}}\ }%
  \textbf{\bibinfo {volume} {18}},\ \bibinfo {pages} {1144} (\bibinfo {year}
  {1978}).%
  \bibAnnoteFile{NoStop}{Ho1978.PRB18.1144}%
\bibitem{Miz2002.PRL89.030401}%
  \BibitemOpen
  \bibfield{author}{%
  \bibinfo {author} {\bibfnamefont{T.}~\bibnamefont{Mizushima}}, \bibinfo
  {author} {\bibfnamefont{K.}~\bibnamefont{Machida}},\ and\ \bibinfo {author}
  {\bibfnamefont{T.}~\bibnamefont{Kita}},\ }%
  \bibfield{journal}{%
  \Doi{10.1103/PhysRevLett.89.030401}{\bibinfo {journal} {Phys. Rev. Lett.}}\
  }%
  \textbf{\bibinfo {volume} {89}},\ \bibinfo {pages} {030401} (\bibinfo {year}
  {2002}).%
  \bibAnnoteFile{NoStop}{Miz2002.PRL89.030401}%
\bibitem{Miz2002.PRA66.053610}%
  \BibitemOpen
  \bibfield{author}{%
  \bibinfo {author} {\bibfnamefont{T.}~\bibnamefont{Mizushima}}, \bibinfo
  {author} {\bibfnamefont{K.}~\bibnamefont{Machida}},\ and\ \bibinfo {author}
  {\bibfnamefont{T.}~\bibnamefont{Kita}},\ }%
  \bibfield{journal}{%
  \Doi{10.1103/PhysRevA.66.053610}{\bibinfo {journal} {Phys. Rev. A}}\ }%
  \textbf{\bibinfo {volume} {66}},\ \bibinfo {pages} {053610} (\bibinfo {year}
  {2002}).%
  \bibAnnoteFile{NoStop}{Miz2002.PRA66.053610}%
\bibitem{Miz2004.PRA70.043613}%
  \BibitemOpen
  \bibfield{author}{%
  \bibinfo {author} {\bibfnamefont{T.}~\bibnamefont{Mizushima}}, \bibinfo
  {author} {\bibfnamefont{N.}~\bibnamefont{Kobayashi}},\ and\ \bibinfo {author}
  {\bibfnamefont{K.}~\bibnamefont{Machida}},\ }%
  \bibfield{journal}{%
  \Doi{10.1103/PhysRevA.70.043613}{\bibinfo {journal} {Phys. Rev. A}}\ }%
  \textbf{\bibinfo {volume} {70}},\ \bibinfo {pages} {043613} (\bibinfo {year}
  {2004}).%
  \bibAnnoteFile{NoStop}{Miz2004.PRA70.043613}%
\bibitem{Dod1997.PRA56.587}%
  \BibitemOpen
  \bibfield{author}{%
  \bibinfo {author} {\bibfnamefont{R.~J.}\ \bibnamefont{Dodd}}, \bibinfo
  {author} {\bibfnamefont{K.}~\bibnamefont{Burnett}}, \bibinfo {author}
  {\bibfnamefont{M.}~\bibnamefont{Edwards}},\ and\ \bibinfo {author}
  {\bibfnamefont{C.~W.}~\bibnamefont{Clark}},\ }%
  \bibfield{journal}{%
  \Doi{10.1103/PhysRevA.56.587}{\bibinfo {journal} {Phys. Rev. A}}\ }%
  \textbf{\bibinfo {volume} {56}},\ \bibinfo {pages} {587} (\bibinfo {year}
  {1997}).%
  \bibAnnoteFile{NoStop}{Dod1997.PRA56.587}%
\bibitem{Rok1997.PRL79.2164}%
  \BibitemOpen
  \bibfield{author}{%
  \bibinfo {author} {\bibfnamefont{D.~S.}\ \bibnamefont{Rokhsar}},\ }%
  \bibfield{journal}{%
  \Doi{10.1103/PhysRevLett.79.2164}{\bibinfo {journal} {Phys. Rev. Lett.}}\ }%
  \textbf{\bibinfo {volume} {79}},\ \bibinfo {pages} {2164} (\bibinfo {year}
  {1997}).%
  \bibAnnoteFile{NoStop}{Rok1997.PRL79.2164}%
\bibitem{Pu1999.PRA59.1533}%
  \BibitemOpen
  \bibfield{author}{%
  \bibinfo {author} {\bibfnamefont{H.}~\bibnamefont{Pu}}, \bibinfo {author}
  {\bibfnamefont{C.~K.}\ \bibnamefont{Law}}, \bibinfo {author}
  {\bibfnamefont{J.~H.}\ \bibnamefont{Eberly}},\ and\ \bibinfo {author}
  {\bibfnamefont{N.~P.}\ \bibnamefont{Bigelow}},\ }%
  \bibfield{journal}{%
  \Doi{10.1103/PhysRevA.59.1533}{\bibinfo {journal} {Phys. Rev. A}}\ }%
  \textbf{\bibinfo {volume} {59}},\ \bibinfo {pages} {1533} (\bibinfo {year}
  {1999}).%
  \bibAnnoteFile{NoStop}{Pu1999.PRA59.1533}%
\bibitem{Iso1999.PRA60.3313}%
  \BibitemOpen
  \bibfield{author}{%
  \bibinfo {author} {\bibfnamefont{T.}~\bibnamefont{Isoshima}}\ and\ \bibinfo
  {author} {\bibfnamefont{K.}~\bibnamefont{Machida}},\ }%
  \bibfield{journal}{%
  \Doi{10.1103/PhysRevA.60.3313}{\bibinfo {journal} {Phys. Rev. A}}\ }%
  \textbf{\bibinfo {volume} {60}},\ \bibinfo {pages} {3313} (\bibinfo {year}
  {1999}).%
  \bibAnnoteFile{NoStop}{Iso1999.PRA60.3313}%
\bibitem{Svi2000.PRL84.5919}%
  \BibitemOpen
  \bibfield{author}{%
  \bibinfo {author} {\bibfnamefont{A.~A.}\ \bibnamefont{Svidzinsky}}\ and\
  \bibinfo {author} {\bibfnamefont{A.~L.}\ \bibnamefont{Fetter}},\ }%
  \bibfield{journal}{%
  \Doi{10.1103/PhysRevLett.84.5919}{\bibinfo {journal} {Phys. Rev. Lett.}}\ }%
  \textbf{\bibinfo {volume} {84}},\ \bibinfo {pages} {5919} (\bibinfo {year}
  {2000}).%
  \bibAnnoteFile{NoStop}{Svi2000.PRL84.5919}%
\bibitem{Vir2001.PRL86.2704}%
  \BibitemOpen
  \bibfield{author}{%
  \bibinfo {author} {\bibfnamefont{S.~M.~M.}\ \bibnamefont{Virtanen}}, \bibinfo
  {author} {\bibfnamefont{T.~P.}\ \bibnamefont{Simula}},\ and\ \bibinfo
  {author} {\bibfnamefont{M.~M.}\ \bibnamefont{Salomaa}},\ }%
  \bibfield{journal}{%
  \Doi{10.1103/PhysRevLett.86.2704}{\bibinfo {journal} {Phys. Rev. Lett.}}\ }%
  \textbf{\bibinfo {volume} {86}},\ \bibinfo {pages} {2704} (\bibinfo {year}
  {2001}).%
  \bibAnnoteFile{NoStop}{Vir2001.PRL86.2704}%
\bibitem{Kaw2004.PRA70.043610}%
  \BibitemOpen
  \bibfield{author}{%
  \bibinfo {author} {\bibfnamefont{Y.}~\bibnamefont{Kawaguchi}}\ and\ \bibinfo
  {author} {\bibfnamefont{T.}~\bibnamefont{Ohmi}},\ }%
  \bibfield{journal}{%
  \Doi{10.1103/PhysRevA.70.043610}{\bibinfo {journal} {Phys. Rev. A}}\ }%
  \textbf{\bibinfo {volume} {70}},\ \bibinfo {pages} {043610} (\bibinfo {year}
  {2004}).%
  \bibAnnoteFile{NoStop}{Kaw2004.PRA70.043610}%
\bibitem{Jac2005.PRA72.053617}%
  \BibitemOpen
  \bibfield{author}{%
  \bibinfo {author} {\bibfnamefont{A.~D.}\ \bibnamefont{Jackson}}, \bibinfo
  {author} {\bibfnamefont{G.~M.}\ \bibnamefont{Kavoulakis}},\ and\ \bibinfo
  {author} {\bibfnamefont{E.}~\bibnamefont{Lundh}},\ }%
  \bibfield{journal}{%
  \Doi{10.1103/PhysRevA.72.053617}{\bibinfo {journal} {Phys. Rev. A}}\ }%
  \textbf{\bibinfo {volume} {72}},\ \bibinfo {pages} {053617} (\bibinfo {year}
  {2005}).%
  \bibAnnoteFile{NoStop}{Jac2005.PRA72.053617}%
\bibitem{Huh2006.PRA74.063619}%
  \BibitemOpen
  \bibfield{author}{%
  \bibinfo {author} {\bibfnamefont{J.~A.~M.}\ \bibnamefont{Huhtam\"aki}},
  \bibinfo {author} {\bibfnamefont{M.}~\bibnamefont{M\"ott\"onen}},\ and\
  \bibinfo {author} {\bibfnamefont{S.~M.~M.}\ \bibnamefont{Virtanen}},\ }%
  \bibfield{journal}{%
  \Doi{10.1103/PhysRevA.74.063619}{\bibinfo {journal} {Phys. Rev. A}}\ }%
  \textbf{\bibinfo {volume} {74}},\ \bibinfo {pages} {063619} (\bibinfo {year}
  {2006}).%
  \bibAnnoteFile{NoStop}{Huh2006.PRA74.063619}%
\bibitem{Lun2006.PRA74.063620}%
  \BibitemOpen
  \bibfield{author}{%
  \bibinfo {author} {\bibfnamefont{E.}~\bibnamefont{Lundh}}\ and\ \bibinfo
  {author} {\bibfnamefont{H.~M.}\ \bibnamefont{Nilsen}},\ }%
  \bibfield{journal}{%
  \Doi{10.1103/PhysRevA.74.063620}{\bibinfo {journal} {Phys. Rev. A}}\ }%
  \textbf{\bibinfo {volume} {74}},\ \bibinfo {pages} {063620} (\bibinfo {year}
  {2006}).%
  \bibAnnoteFile{NoStop}{Lun2006.PRA74.063620}%
\bibitem{Cap2009.JPB42.145301}%
  \BibitemOpen
  \bibfield{author}{%
  \bibinfo {author} {\bibfnamefont{P.}~\bibnamefont{Capuzzi}}\ and\ \bibinfo
  {author} {\bibfnamefont{D.~M.}\ \bibnamefont{Jezek}},\ }%
  \bibfield{journal}{%
  \Doi{10.1088/0953-4075/42/14/145301}{\bibinfo {journal} {J. Phys. B: At. Mol.
  Opt. Phys.}}\ }%
  \textbf{\bibinfo {volume} {42}},\ \bibinfo {pages} {145301} (\bibinfo {year}
  {2009}).%
  \bibAnnoteFile{NoStop}{Cap2009.JPB42.145301}%
\bibitem{Kuo2010.PRA81.023603}%
  \BibitemOpen
  \bibfield{author}{%
  \bibinfo {author} {\bibfnamefont{P.}~\bibnamefont{Kuopanportti}}, \bibinfo
  {author} {\bibfnamefont{E.}~\bibnamefont{Lundh}}, \bibinfo {author}
  {\bibfnamefont{J.~A.~M.}\ \bibnamefont{Huhtam\"aki}}, \bibinfo {author}
  {\bibfnamefont{V.}~\bibnamefont{Pietil\"a}},\ and\ \bibinfo {author}
  {\bibfnamefont{M.}~\bibnamefont{M\"ott\"onen}},\ }%
  \bibfield{journal}{%
  \Doi{10.1103/PhysRevA.81.023603}{\bibinfo {journal} {Phys. Rev. A}}\ }%
  \textbf{\bibinfo {volume} {81}},\ \bibinfo {pages} {023603} (\bibinfo {year}
  {2010}).%
  \bibAnnoteFile{NoStop}{Kuo2010.PRA81.023603}%
\bibitem{Kuo2010.PRA81.033627}%
  \BibitemOpen
  \bibfield{author}{%
  \bibinfo {author} {\bibfnamefont{P.}~\bibnamefont{Kuopanportti}}\ and\
  \bibinfo {author} {\bibfnamefont{M.}~\bibnamefont{M\"ott\"onen}},\ }%
  \bibfield{journal}{%
  \Doi{10.1103/PhysRevA.81.033627}{\bibinfo {journal} {Phys. Rev. A}}\ }%
  \textbf{\bibinfo {volume} {81}},\ \bibinfo {pages} {033627} (\bibinfo {year}
  {2010}).%
  \bibAnnoteFile{NoStop}{Kuo2010.PRA81.033627}%
\bibitem{Chi2010.RMP82.1225}%
  \BibitemOpen
  \bibfield{author}{%
  \bibinfo {author} {\bibfnamefont{C.}~\bibnamefont{Chin}}, \bibinfo {author}
  {\bibfnamefont{R.}~\bibnamefont{Grimm}}, \bibinfo {author}
  {\bibfnamefont{P.}~\bibnamefont{Julienne}},\ and\ \bibinfo {author}
  {\bibfnamefont{E.}~\bibnamefont{Tiesinga}},\ }%
  \bibfield{journal}{%
  \Doi{10.1103/RevModPhys.82.1225}{\bibinfo {journal} {Rev. Mod. Phys.}}\ }%
  \textbf{\bibinfo {volume} {82}},\ \bibinfo {pages} {1225} (\bibinfo {year}
  {2010}).%
  \bibAnnoteFile{NoStop}{Chi2010.RMP82.1225}%
\bibitem{Mar2002.PRL89.283202}%
  \BibitemOpen
  \bibfield{author}{%
  \bibinfo {author} {\bibfnamefont{A.}~\bibnamefont{Marte}}, \bibinfo {author}
  {\bibfnamefont{T.}~\bibnamefont{Volz}}, \bibinfo {author}
  {\bibfnamefont{J.}~\bibnamefont{Schuster}}, \bibinfo {author}
  {\bibfnamefont{S.}~\bibnamefont{D\"urr}}, \bibinfo {author}
  {\bibfnamefont{G.}~\bibnamefont{Rempe}}, \bibinfo {author}
  {\bibfnamefont{E.~G.~M.}\ \bibnamefont{van Kempen}},\ and\ \bibinfo {author}
  {\bibfnamefont{B.~J.}\ \bibnamefont{Verhaar}},\ }%
  \bibfield{journal}{%
  \Doi{10.1103/PhysRevLett.89.283202}{\bibinfo {journal} {Phys. Rev. Lett.}}\
  }%
  \textbf{\bibinfo {volume} {89}},\ \bibinfo {pages} {283202} (\bibinfo {year}
  {2002}).%
  \bibAnnoteFile{NoStop}{Mar2002.PRL89.283202}%
\bibitem{Er2004.PRA69.032705}%
  \BibitemOpen
  \bibfield{author}{%
  \bibinfo {author} {\bibfnamefont{M.}~\bibnamefont{Erhard}}, \bibinfo {author}
  {\bibfnamefont{H.}~\bibnamefont{Schmaljohann}}, \bibinfo {author}
  {\bibfnamefont{J.}~\bibnamefont{Kronj\"ager}}, \bibinfo {author}
  {\bibfnamefont{K.}~\bibnamefont{Bongs}},\ and\ \bibinfo {author}
  {\bibfnamefont{K.}~\bibnamefont{Sengstock}},\ }%
  \bibfield{journal}{%
  \Doi{10.1103/PhysRevA.69.032705}{\bibinfo {journal} {Phys. Rev. A}}\ }%
  \textbf{\bibinfo {volume} {69}},\ \bibinfo {pages} {032705} (\bibinfo {year}
  {2004}).%
  \bibAnnoteFile{NoStop}{Er2004.PRA69.032705}%
\bibitem{Wid2004.PRL92.160406}%
  \BibitemOpen
  \bibfield{author}{%
  \bibinfo {author} {\bibfnamefont{A.}~\bibnamefont{Widera}}, \bibinfo {author}
  {\bibfnamefont{O.}~\bibnamefont{Mandel}}, \bibinfo {author}
  {\bibfnamefont{M.}~\bibnamefont{Greiner}}, \bibinfo {author}
  {\bibfnamefont{S.}~\bibnamefont{Kreim}}, \bibinfo {author}
  {\bibfnamefont{T.~W.}\ \bibnamefont{H\"ansch}},\ and\ \bibinfo {author}
  {\bibfnamefont{I.}~\bibnamefont{Bloch}},\ }%
  \bibfield{journal}{%
  \Doi{10.1103/PhysRevLett.92.160406}{\bibinfo {journal} {Phys. Rev. Lett.}}\
  }%
  \textbf{\bibinfo {volume} {92}},\ \bibinfo {pages} {160406} (\bibinfo {year}
  {2004}).%
  \bibAnnoteFile{NoStop}{Wid2004.PRL92.160406}%
\bibitem{Err2007.NJP9.223}%
  \BibitemOpen
  \bibfield{author}{%
  \bibinfo {author} {\bibfnamefont{C.}~\bibnamefont{D'Errico}}, \bibinfo
  {author} {\bibfnamefont{M.}~\bibnamefont{Zaccanti}}, \bibinfo {author}
  {\bibfnamefont{M.}~\bibnamefont{Fattori}}, \bibinfo {author}
  {\bibfnamefont{G.}~\bibnamefont{Roati}}, \bibinfo {author}
  {\bibfnamefont{M.}~\bibnamefont{Inguscio}}, \bibinfo {author}
  {\bibfnamefont{G.}~\bibnamefont{Modugno}},\ and\ \bibinfo {author}
  {\bibfnamefont{A.}~\bibnamefont{Simoni}},\ }%
  \bibfield{journal}{%
  \Doi{10.1088/1367-2630/9/7/223}{\bibinfo {journal} {New J. Phys.}}\ }%
  \textbf{\bibinfo {volume} {9}},\ \bibinfo {pages} {223} (\bibinfo {year}
  {2007}).%
  \bibAnnoteFile{NoStop}{Err2007.NJP9.223}%
\bibitem{Kis2009.PRA79.031602}%
  \BibitemOpen
  \bibfield{author}{%
  \bibinfo {author} {\bibfnamefont{T.}~\bibnamefont{Kishimoto}}, \bibinfo
  {author} {\bibfnamefont{J.}~\bibnamefont{Kobayashi}}, \bibinfo {author}
  {\bibfnamefont{K.}~\bibnamefont{Noda}}, \bibinfo {author}
  {\bibfnamefont{K.}~\bibnamefont{Aikawa}}, \bibinfo {author}
  {\bibfnamefont{M.}~\bibnamefont{Ueda}},\ and\ \bibinfo {author}
  {\bibfnamefont{S.}~\bibnamefont{Inouye}},\ }%
  \bibfield{journal}{%
  \Doi{10.1103/PhysRevA.79.031602}{\bibinfo {journal} {Phys. Rev. A}}\ }%
  \textbf{\bibinfo {volume} {79}},\ \bibinfo {pages} {031602} (\bibinfo {year}
  {2009}).%
  \bibAnnoteFile{NoStop}{Kis2009.PRA79.031602}%
\bibitem{Kem2002.PRL88.093201}%
  \BibitemOpen
  \bibfield{author}{%
  \bibinfo {author} {\bibfnamefont{E.~G.~M.}\ \bibnamefont{van Kempen}},
  \bibinfo {author} {\bibfnamefont{S.~J. J. M.~F.}\ \bibnamefont{Kokkelmans}},
  \bibinfo {author} {\bibfnamefont{D.~J.}\ \bibnamefont{Heinzen}},\ and\
  \bibinfo {author} {\bibfnamefont{B.~J.}\ \bibnamefont{Verhaar}},\ }%
  \bibfield{journal}{%
  \Doi{10.1103/PhysRevLett.88.093201}{\bibinfo {journal} {Phys. Rev. Lett.}}\
  }%
  \textbf{\bibinfo {volume} {88}},\ \bibinfo {pages} {093201} (\bibinfo {year}
  {2002}).%
  \bibAnnoteFile{NoStop}{Kem2002.PRL88.093201}%
\bibitem{Wan2000.PRA62.052704}%
  \BibitemOpen
  \bibfield{author}{%
  \bibinfo {author} {\bibfnamefont{H.}~\bibnamefont{Wang}}, \bibinfo {author}
  {\bibfnamefont{A.~N.}\ \bibnamefont{Nikolov}}, \bibinfo {author}
  {\bibfnamefont{J.~R.}\ \bibnamefont{Ensher}}, \bibinfo {author}
  {\bibfnamefont{P.~L.}\ \bibnamefont{Gould}}, \bibinfo {author}
  {\bibfnamefont{E.~E.}\ \bibnamefont{Eyler}}, \bibinfo {author}
  {\bibfnamefont{W.~C.}\ \bibnamefont{Stwalley}}, \bibinfo {author}
  {\bibfnamefont{J.~P.}\ \bibnamefont{Burke}}, \bibinfo {author}
  {\bibfnamefont{J.~L.}\ \bibnamefont{Bohn}}, \bibinfo {author}
  {\bibfnamefont{C.~H.}\ \bibnamefont{Greene}}, \bibinfo {author}
  {\bibfnamefont{E.}~\bibnamefont{Tiesinga}}, \bibinfo {author}
  {\bibfnamefont{C.~J.}\ \bibnamefont{Williams}},\ and\ \bibinfo {author}
  {\bibfnamefont{P.~S.}\ \bibnamefont{Julienne}},\ }%
  \bibfield{journal}{%
  \Doi{10.1103/PhysRevA.62.052704}{\bibinfo {journal} {Phys. Rev. A}}\ }%
  \textbf{\bibinfo {volume} {62}},\ \bibinfo {pages} {052704} (\bibinfo {year}
  {2000}).%
  \bibAnnoteFile{NoStop}{Wan2000.PRA62.052704}%
\bibitem{Fis2003.PRL90.140402}%
  \BibitemOpen
  \bibfield{author}{%
  \bibinfo {author} {\bibfnamefont{U.~R.}\ \bibnamefont{Fischer}}\ and\
  \bibinfo {author} {\bibfnamefont{G.}~\bibnamefont{Baym}},\ }%
  \bibfield{journal}{%
  \Doi{10.1103/PhysRevLett.90.140402}{\bibinfo {journal} {Phys. Rev. Lett.}}\
  }%
  \textbf{\bibinfo {volume} {90}},\ \bibinfo {pages} {140402} (\bibinfo {year}
  {2003}).%
  \bibAnnoteFile{NoStop}{Fis2003.PRL90.140402}%
\bibitem{Jos2004.Chaos14.875}%
  \BibitemOpen
  \bibfield{author}{%
  \bibinfo {author} {\bibfnamefont{C.}~\bibnamefont{Josserand}},\ }%
  \bibfield{journal}{%
  \Doi{http://dx.doi.org/10.1063/1.1785892}{\bibinfo {journal} {Chaos}}\ }%
  \textbf{\bibinfo {volume} {14}},\ \bibinfo {pages} {875} (\bibinfo {year}
  {2004}).%
  \bibAnnoteFile{NoStop}{Jos2004.Chaos14.875}%
\bibitem{Aft2004.PRA69.033608}%
  \BibitemOpen
  \bibfield{author}{%
  \bibinfo {author} {\bibfnamefont{A.}~\bibnamefont{Aftalion}}\ and\ \bibinfo
  {author} {\bibfnamefont{I.}~\bibnamefont{Danaila}},\ }%
  \bibfield{journal}{%
  \Doi{10.1103/PhysRevA.69.033608}{\bibinfo {journal} {Phys. Rev. A}}\ }%
  \textbf{\bibinfo {volume} {69}},\ \bibinfo {pages} {033608} (\bibinfo {year}
  {2004}).%
  \bibAnnoteFile{NoStop}{Aft2004.PRA69.033608}%
\bibitem{Jac2004.PRA69.053619}%
  \BibitemOpen
  \bibfield{author}{%
  \bibinfo {author} {\bibfnamefont{A.~D.}\ \bibnamefont{Jackson}}, \bibinfo
  {author} {\bibfnamefont{G.~M.}\ \bibnamefont{Kavoulakis}},\ and\ \bibinfo
  {author} {\bibfnamefont{E.}~\bibnamefont{Lundh}},\ }%
  \bibfield{journal}{%
  \Doi{10.1103/PhysRevA.69.053619}{\bibinfo {journal} {Phys. Rev. A}}\ }%
  \textbf{\bibinfo {volume} {69}},\ \bibinfo {pages} {053619} (\bibinfo {year}
  {2004}).%
  \bibAnnoteFile{NoStop}{Jac2004.PRA69.053619}%
\bibitem{Fet2005.PRA71.013605}%
  \BibitemOpen
  \bibfield{author}{%
  \bibinfo {author} {\bibfnamefont{A.~L.}\ \bibnamefont{Fetter}}, \bibinfo
  {author} {\bibfnamefont{B.}~\bibnamefont{Jackson}},\ and\ \bibinfo {author}
  {\bibfnamefont{S.}~\bibnamefont{Stringari}},\ }%
  \bibfield{journal}{%
  \Doi{10.1103/PhysRevA.71.013605}{\bibinfo {journal} {Phys. Rev. A}}\ }%
  \textbf{\bibinfo {volume} {71}},\ \bibinfo {pages} {013605} (\bibinfo {year}
  {2005}).%
  \bibAnnoteFile{NoStop}{Fet2005.PRA71.013605}%
\bibitem{Kim2005.PRA72.023619}%
  \BibitemOpen
  \bibfield{author}{%
  \bibinfo {author} {\bibfnamefont{J.-k.}\ \bibnamefont{Kim}}\ and\ \bibinfo
  {author} {\bibfnamefont{A.~L.}\ \bibnamefont{Fetter}},\ }%
  \bibfield{journal}{%
  \Doi{10.1103/PhysRevA.72.023619}{\bibinfo {journal} {Phys. Rev. A}}\ }%
  \textbf{\bibinfo {volume} {72}},\ \bibinfo {pages} {023619} (\bibinfo {year}
  {2005}).%
  \bibAnnoteFile{NoStop}{Kim2005.PRA72.023619}%
\bibitem{Fu2006.PRA73.013614}%
  \BibitemOpen
  \bibfield{author}{%
  \bibinfo {author} {\bibfnamefont{H.}~\bibnamefont{Fu}}\ and\ \bibinfo
  {author} {\bibfnamefont{E.}~\bibnamefont{Zaremba}},\ }%
  \bibfield{journal}{%
  \Doi{10.1103/PhysRevA.73.013614}{\bibinfo {journal} {Phys. Rev. A}}\ }%
  \textbf{\bibinfo {volume} {73}},\ \bibinfo {pages} {013614} (\bibinfo {year}
  {2006}).%
  \bibAnnoteFile{NoStop}{Fu2006.PRA73.013614}%
\bibitem{Cor2007.JMP48.042104}%
  \BibitemOpen
  \bibfield{author}{%
  \bibinfo {author} {\bibfnamefont{M.}~\bibnamefont{Correggi}}, \bibinfo
  {author} {\bibfnamefont{T.}~\bibnamefont{Rindler-Daller}},\ and\ \bibinfo
  {author} {\bibfnamefont{J.}~\bibnamefont{Yngvason}},\ }%
  \bibfield{journal}{%
  \Doi{http://dx.doi.org/10.1063/1.2712421}{\bibinfo {journal} {J. Math.
  Phys.}}\ }%
  \textbf{\bibinfo {volume} {48}},\ \bibinfo {pages} {042104} (\bibinfo {year}
  {2007}).%
  \bibAnnoteFile{NoStop}{Cor2007.JMP48.042104}%
\bibitem{Cor2011.PRA84.053614}%
  \BibitemOpen
  \bibfield{author}{%
  \bibinfo {author} {\bibfnamefont{M.}~\bibnamefont{Correggi}}, \bibinfo
  {author} {\bibfnamefont{F.}~\bibnamefont{Pinsker}}, \bibinfo {author}
  {\bibfnamefont{N.}~\bibnamefont{Rougerie}},\ and\ \bibinfo {author}
  {\bibfnamefont{J.}~\bibnamefont{Yngvason}},\ }%
  \bibfield{journal}{%
  \Doi{10.1103/PhysRevA.84.053614}{\bibinfo {journal} {Phys. Rev. A}}\ }%
  \textbf{\bibinfo {volume} {84}},\ \bibinfo {pages} {053614} (\bibinfo {year}
  {2011}).%
  \bibAnnoteFile{NoStop}{Cor2011.PRA84.053614}%
\bibitem{Mat1999.PRL83.3358}%
  \BibitemOpen
  \bibfield{author}{%
  \bibinfo {author} {\bibfnamefont{M.~R.}\ \bibnamefont{Matthews}}, \bibinfo
  {author} {\bibfnamefont{B.~P.}\ \bibnamefont{Anderson}}, \bibinfo {author}
  {\bibfnamefont{P.~C.}\ \bibnamefont{Haljan}}, \bibinfo {author}
  {\bibfnamefont{D.~S.}\ \bibnamefont{Hall}}, \bibinfo {author}
  {\bibfnamefont{M.~J.}\ \bibnamefont{Holland}}, \bibinfo {author}
  {\bibfnamefont{J.~E.}\ \bibnamefont{Williams}}, \bibinfo {author}
  {\bibfnamefont{C.~E.}\ \bibnamefont{Wieman}},\ and\ \bibinfo {author}
  {\bibfnamefont{E.~A.}\ \bibnamefont{Cornell}},\ }%
  \bibfield{journal}{%
  \Doi{10.1103/PhysRevLett.83.3358}{\bibinfo {journal} {Phys. Rev. Lett.}}\ }%
  \textbf{\bibinfo {volume} {83}},\ \bibinfo {pages} {3358} (\bibinfo {year}
  {1999}).%
  \bibAnnoteFile{NoStop}{Mat1999.PRL83.3358}%
\bibitem{Mue2004.PRA69.033606}%
  \BibitemOpen
  \bibfield{author}{%
  \bibinfo {author} {\bibfnamefont{E.~J.}\ \bibnamefont{Mueller}},\ }%
  \bibfield{journal}{%
  \Doi{10.1103/PhysRevA.69.033606}{\bibinfo {journal} {Phys. Rev. A}}\ }%
  \textbf{\bibinfo {volume} {69}},\ \bibinfo {pages} {033606} (\bibinfo {year}
  {2004}).%
  \bibAnnoteFile{NoStop}{Mue2004.PRA69.033606}%
\bibitem{Mak1977.JLTP27.635}%
  \BibitemOpen
  \bibfield{author}{%
  \bibinfo {author} {\bibfnamefont{K.}~\bibnamefont{Maki}}\ and\ \bibinfo
  {author} {\bibfnamefont{T.}~\bibnamefont{Tsuneto}},\ }%
  \bibfield{journal}{%
  \Doi{10.1007/BF00655292}{\bibinfo {journal} {J. Low Temp. Phys.}}\ }%
  \textbf{\bibinfo {volume} {27}},\ \bibinfo {pages} {635} (\bibinfo {year}
  {1977}).%
  \bibAnnoteFile{NoStop}{Mak1977.JLTP27.635}%
\bibitem{Mer1979.RMP51.591}%
  \BibitemOpen
  \bibfield{author}{%
  \bibinfo {author} {\bibfnamefont{N.~D.}\ \bibnamefont{Mermin}},\ }%
  \bibfield{journal}{%
  \Doi{10.1103/RevModPhys.51.591}{\bibinfo {journal} {Rev. Mod. Phys.}}\ }%
  \textbf{\bibinfo {volume} {51}},\ \bibinfo {pages} {591} (\bibinfo {year}
  {1979}).%
  \bibAnnoteFile{NoStop}{Mer1979.RMP51.591}%
\bibitem{Sim2002.PRA65.033614}%
  \BibitemOpen
  \bibfield{author}{%
  \bibinfo {author} {\bibfnamefont{T.~P.}\ \bibnamefont{Simula}}, \bibinfo
  {author} {\bibfnamefont{S.~M.~M.}\ \bibnamefont{Virtanen}},\ and\ \bibinfo
  {author} {\bibfnamefont{M.~M.}\ \bibnamefont{Salomaa}},\ }%
  \bibfield{journal}{%
  \Doi{10.1103/PhysRevA.65.033614}{\bibinfo {journal} {Phys. Rev. A}}\ }%
  \textbf{\bibinfo {volume} {65}},\ \bibinfo {pages} {033614} (\bibinfo {year}
  {2002}).%
  \bibAnnoteFile{NoStop}{Sim2002.PRA65.033614}%
\bibitem{Kuo2010.JLTP161.561}%
  \BibitemOpen
  \bibfield{author}{%
  \bibinfo {author} {\bibfnamefont{P.}~\bibnamefont{Kuopanportti}}\ and\
  \bibinfo {author} {\bibfnamefont{M.}~\bibnamefont{M\"ott\"onen}},\ }%
  \bibfield{journal}{%
  \Doi{10.1007/s10909-010-0216-1}{\bibinfo {journal} {J. Low Temp. Phys.}}\ }%
  \textbf{\bibinfo {volume} {161}},\ \bibinfo {pages} {561} (\bibinfo {year}
  {2010}).%
  \bibAnnoteFile{NoStop}{Kuo2010.JLTP161.561}%
\bibitem{Kar2013.PRA87.043609}%
  \BibitemOpen
  \bibfield{author}{%
  \bibinfo {author} {\bibfnamefont{E.~O.}\ \bibnamefont{Karabulut}}, \bibinfo
  {author} {\bibfnamefont{F.}~\bibnamefont{Malet}}, \bibinfo {author}
  {\bibfnamefont{G.~M.}\ \bibnamefont{Kavoulakis}},\ and\ \bibinfo {author}
  {\bibfnamefont{S.~M.}\ \bibnamefont{Reimann}},\ }%
  \bibfield{journal}{%
  \Doi{10.1103/PhysRevA.87.043609}{\bibinfo {journal} {Phys. Rev. A}}\ }%
  \textbf{\bibinfo {volume} {87}},\ \bibinfo {pages} {043609} (\bibinfo {year}
  {2013}).%
  \bibAnnoteFile{NoStop}{Kar2013.PRA87.043609}%
\bibitem{Che2000.PRL85.2223}%
  \BibitemOpen
  \bibfield{author}{%
  \bibinfo {author} {\bibfnamefont{F.}~\bibnamefont{Chevy}}, \bibinfo {author}
  {\bibfnamefont{K.~W.}\ \bibnamefont{Madison}},\ and\ \bibinfo {author}
  {\bibfnamefont{J.}~\bibnamefont{Dalibard}},\ }%
  \bibfield{journal}{%
  \Doi{10.1103/PhysRevLett.85.2223}{\bibinfo {journal} {Phys. Rev. Lett.}}\ }%
  \textbf{\bibinfo {volume} {85}},\ \bibinfo {pages} {2223} (\bibinfo {year}
  {2000}).%
  \bibAnnoteFile{NoStop}{Che2000.PRL85.2223}%
\bibitem{Hal2001.PRL86.2922}%
  \BibitemOpen
  \bibfield{author}{%
  \bibinfo {author} {\bibfnamefont{P.~C.}\ \bibnamefont{Haljan}}, \bibinfo
  {author} {\bibfnamefont{B.~P.}\ \bibnamefont{Anderson}}, \bibinfo {author}
  {\bibfnamefont{I.}~\bibnamefont{Coddington}},\ and\ \bibinfo {author}
  {\bibfnamefont{E.~A.}\ \bibnamefont{Cornell}},\ }%
  \bibfield{journal}{%
  \Doi{10.1103/PhysRevLett.86.2922}{\bibinfo {journal} {Phys. Rev. Lett.}}\ }%
  \textbf{\bibinfo {volume} {86}},\ \bibinfo {pages} {2922} (\bibinfo {year}
  {2001}).%
  \bibAnnoteFile{NoStop}{Hal2001.PRL86.2922}%
\bibitem{Lea2002.PRL89.190403}%
  \BibitemOpen
  \bibfield{author}{%
  \bibinfo {author} {\bibfnamefont{A.~E.}\ \bibnamefont{Leanhardt}}, \bibinfo
  {author} {\bibfnamefont{A.}~\bibnamefont{G\"orlitz}}, \bibinfo {author}
  {\bibfnamefont{A.~P.}\ \bibnamefont{Chikkatur}}, \bibinfo {author}
  {\bibfnamefont{D.}~\bibnamefont{Kielpinski}}, \bibinfo {author}
  {\bibfnamefont{Y.-i.}\ \bibnamefont{Shin}}, \bibinfo {author}
  {\bibfnamefont{D.~E.}\ \bibnamefont{Pritchard}},\ and\ \bibinfo {author}
  {\bibfnamefont{W.}~\bibnamefont{Ketterle}},\ }%
  \bibfield{journal}{%
  \Doi{10.1103/PhysRevLett.89.190403}{\bibinfo {journal} {Phys. Rev. Lett.}}\
  }%
  \textbf{\bibinfo {volume} {89}},\ \bibinfo {pages} {190403} (\bibinfo {year}
  {2002}).%
  \bibAnnoteFile{NoStop}{Lea2002.PRL89.190403}%
\bibitem{Seo2014.JKPS64.53}%
  \BibitemOpen
  \bibfield{author}{%
  \bibinfo {author} {\bibfnamefont{S.}~\bibnamefont{Seo}}, \bibinfo {author}
  {\bibfnamefont{J.-y.}\ \bibnamefont{Choi}},\ and\ \bibinfo {author}
  {\bibfnamefont{Y.-i.}\ \bibnamefont{Shin}},\ }%
  \bibfield{journal}{%
  \Doi{10.3938/jkps.64.53}{\bibinfo {journal} {J. Korean Phys. Soc.}}\ }%
  \textbf{\bibinfo {volume} {64}},\ \bibinfo {pages} {53} (\bibinfo {year}
  {2014}).%
  \bibAnnoteFile{NoStop}{Seo2014.JKPS64.53}%
\bibitem{Geu2010.PRB81.214514}%
  \BibitemOpen
  \bibfield{author}{%
  \bibinfo {author} {\bibfnamefont{R.}~\bibnamefont{Geurts}}, \bibinfo {author}
  {\bibfnamefont{M.~V.}\ \bibnamefont{Milo\ifmmode \check{s}\else
  \v{s}\fi{}evi\ifmmode~\acute{c}\else \'{c}\fi{}}},\ and\ \bibinfo {author}
  {\bibfnamefont{F.~M.}\ \bibnamefont{Peeters}},\ }%
  \bibfield{journal}{%
  \Doi{10.1103/PhysRevB.81.214514}{\bibinfo {journal} {Phys. Rev. B}}\ }%
  \textbf{\bibinfo {volume} {81}},\ \bibinfo {pages} {214514} (\bibinfo {year}
  {2010}).%
  \bibAnnoteFile{NoStop}{Geu2010.PRB81.214514}%
\bibitem{Gar2011.PRL107.197001}%
  \BibitemOpen
  \bibfield{author}{%
  \bibinfo {author} {\bibfnamefont{J.}~\bibnamefont{Garaud}}, \bibinfo {author}
  {\bibfnamefont{J.}~\bibnamefont{Carlstr\"om}},\ and\ \bibinfo {author}
  {\bibfnamefont{E.}~\bibnamefont{Babaev}},\ }%
  \bibfield{journal}{%
  \Doi{10.1103/PhysRevLett.107.197001}{\bibinfo {journal} {Phys. Rev. Lett.}}\
  }%
  \textbf{\bibinfo {volume} {107}},\ \bibinfo {pages} {197001} (\bibinfo {year}
  {2011}).%
  \bibAnnoteFile{NoStop}{Gar2011.PRL107.197001}%
\end{thebibliography}%
\end{document}